\documentclass{article}

\usepackage{PRIMEarxiv}

\usepackage[utf8]{inputenc} 
\usepackage[T1]{fontenc}    
\usepackage{hyperref}       
\usepackage{url}            
\usepackage{booktabs}       
\usepackage{amsfonts}       
\usepackage{amsmath}
\usepackage{nicefrac}       
\usepackage{microtype}      
\usepackage{lipsum}
\usepackage{graphicx}
\usepackage{subcaption}
\usepackage{mathpazo}
\usepackage{sectsty}
\usepackage{tikz}
\usepackage[capitalize]{cleveref}
\usepackage[shortlabels,inline]{enumitem}
\usepackage{float}
\usepackage{natbib}
\usepackage{csquotes}
\usepackage{tabularx}
\newcolumntype{L}{@{\extracolsep{\fill}}l}
\newcolumntype{R}{@{\extracolsep{\fill}}r}
\newcolumntype{C}{@{\extracolsep{\fill}}c}

\title{Data-driven uncertainty-aware seakeeping prediction of the Delft 372 catamaran using ensemble Hankel dynamic mode decomposition}

\author{
  Giorgio Palma$^{a, \star}$, Andrea Serani$^{a}$, Matteo Diez$^{a}$\\
  $^{a}$National Research Council-Institute of Marine Engineering, Via di Vallerano 139, Rome, 00128, Italy\\
  $^\star$\texttt{giorgio.palma@cnr.it} \\
}

\begin{document}

\begin{tikzpicture}[remember picture,overlay]
   \node [rectangle, fill=cyan, fill opacity=0.5, anchor=north, minimum width=\paperwidth, minimum height=3cm] at (current page.north) {};

   \node [anchor=north, minimum width=\paperwidth, minimum height=3cm, text width=\textwidth, align=center, text height=5ex, text depth=15ex, align=left] at (current page.north) {
     \sffamily\small
     \textbf{This is a preprint submitted to:} \textit{Journal of Hydrodynamics}
   };
\end{tikzpicture}

\maketitle

\begin{abstract}
In this study, we present and validate an ensemble‐based Hankel Dynamic Mode Decomposition with control (HDMDc) for uncertainty-aware seakeeping predictions of a high-speed catamaran, namely the Delft 372 model. Experimental measurements (time histories) of wave elevation at the longitudinal center of gravity, heave, pitch, notional flight-deck velocity, notional bridge acceleration, and total resistance were collected from irregular wave basin tests on a 1:33.3 scale replica of the Delft 372 model under sea state 5 conditions at Fr = 0.425, and organized into training, validation, and test sets.
The HDMDc algorithm constructs an equation-free linear reduced-order model of the seakeeping vessel by augmenting states and inputs with their time-lagged copies to capture nonlinear and memory effects. Two ensembling strategies, namely Bayesian HDMDc (BHDMDc), which samples hyperparameters considered stochastic variables with prior distribution to produce posterior mean forecasts with confidence intervals, and Frequentist HDMDc (FHDMDc), which aggregates multiple model obtained over data subsets, are compared in providing seakeeping prediction and uncertainty quantification. The FHDMDc approach is found to improve the accuracy of the predictions compared to the deterministic counterpart, also providing robust uncertainty estimation; whereas the application of BHDMDc to the present test case is not found beneficial in comparison to the deterministic model. 
FHDMDc-derived probability density functions for the motions closely match both experimental data and URANS results, demonstrating reliable and computationally efficient seakeeping prediction for design and operational support.
\end{abstract}


\section*{Introduction}
Accurate prediction of seakeeping performance is critical in the design and operational assessment of marine vessels, for ensuring operational safety, and structural integrity. 
In high-performance marine vessels such as catamarans, the complex hydrodynamic interactions between twin hulls and their unique dynamic behavior in waves pose specific challenges for traditional modeling techniques.

Computational fluid dynamics (CFD) methods, based on (unsteady) Reynolds-averaged Navier-Stokes and large eddy simulations, offer high-fidelity modeling of hydrodynamic forces and flow fields, capable of capturing complex nonlinear fluid-structure interactions. The accuracy of CFD predictions for ship performance has been proven for both monohulls \cite{serani2021urans,aram2024cfd} by comparing simulation results to experimental (EFD) data in extreme wave conditions generating highly non-linear dynamics, and twin hull configurations, addressing high-speed seakeeping in high amplitude waves \cite{castiglione2011}, steady drift \cite{broglia2015cfd, broglia2019}, and water-jet/maneuvering \cite{sadat2013cfd}.
However, their high computational cost can be prohibitive for real-time applications or iterative design processes.

The experimental and computational work conducted in \cite{diez2018statistical} and \cite{DURANTE2020} provides a comprehensive benchmark for the validation of ship response predictions in irregular seas, focusing on uncertainty quantification (UQ) methodologies. Both studies utilize the Delft 372 catamaran model in head waves representative of sea state 5 at a Froude number of 0.425. The experimental campaigns include seakeeping tests in both regular and irregular waves, with high-resolution measurements of axial force, heave and pitch motions, vertical acceleration of a notional bridge, and vertical velocity of a notional flight-deck. In parallel, high-fidelity CFD simulations are performed under matching conditions to enable direct comparisons. Statistical errors and uncertainties in both EFD and CFD are assessed in \cite{diez2018statistical} via autocovariance analysis and bootstrap techniques, proposing a regular wave-based UQ model as a cost-effective alternative for estimating ship response statistics in irregular seas. The regular wave UQ model is validated against experimentally derived statistical estimators such as expected value, standard deviation, and quantiles of both primary and secondary variables. Together, these works establish a validated experimental and numerical framework for UQ in seakeeping performance evaluation, laying the foundation for robust design under realistic ocean conditions.


Equation-based reduced order models (ROMs), such as the Maneuvering Modeling Group (MMG) model \cite{Yasukawa2015,Yasukawa2016}, have been developed as physics-based efficient alternatives, and demonstrated good agreement with experiments and CFD for maneuvering of displacement ships, \cite{Sanada2021,stern2022kcs}, planing hulls \cite{diez2024sname}, and have been recently studied also for twin-hull configurations \cite{pandey2016manoeuvring,PANDEY2016}. 
Despite the promising results, such models typically require a large amount of data of CFD computations for their training and definition of forcing terms.

In recent years, data-driven methods have emerged as a promising alternative for system identification and ship motion prediction. 
Machine learning techniques gained popularity due to their ability to model complex input-output relations in an automated manner directly from data. 
In particular, recurrent neural networks (RNN) and long short-term memory networks (LSTM), along with their bidirectional LSTM (BiLSTM) variant, have been demonstrated effective for building equation-free data-driven models for modeling time-series ship motion data, and provide multi-step ahead forecasting of ship degree-of-freedom in calm water and waves \cite{Diez2024}.
The performance of RNN, LSTM, and gated recurrent unit (GRU) models is compared in \cite{DAgostino2022} as a proof of concept for the forecasting of a self-propelled destroyer-type vessel, sailing in stern-quartering sea state 7, using a sequence-to-sequence modeling. 
LSTM are used in \cite{XU2021} to create a nonlinear model for ship heave and roll using wave elevation as input. 
In \cite{Wang2023}, BiLSTM are used to build a short-term predictor for course angle, yaw rate, roll angle, and total speed using course angle, yaw rate, roll angle, total speed, and rudder angle as inputs.
In \cite{jiang2024}, motion and wave elevation data are used to train several BiLSTM networks for different ship operating conditions, then a dynamic model averaging is employed to interpolate the ship pitch and heave prediction for the test condition.
Uncertainty estimation has been included in an LSTM-based system identification method in \cite{silva2022} for the 6-DoF of a displacement hull in course-keeping and turning circle by using a Monte Carlo dropout approach during prediction. The different models obtained after the dropout provide an ensemble of predictions, from which an average prediction is calculated along with an uncertainty estimate.
The strength of machine and deep learning methods lies in their ability to capture relevant hidden and nonlinear dynamics directly from available data, their compactness, and fast evaluation. 
However, while powerful, such models are often considered black-box approaches and pose challenges in terms of physical interpretability.
In addition, deep learning models typically require large datasets of high-fidelity data for training (more complex architectures usually require more expensive training) and to generalize effectively.

In contrast, Dynamic Mode Decomposition (DMD) offers data-driven equation-free modeling, with minimal or no prior knowledge of the system equations, still ensuring a certain level of interpretability.
Originally introduced in the fluid dynamics community by Schmid \cite{schmid2008dynamic,schmid2010}, DMD approximates nonlinear dynamics through a linear map that best advances the system from one state to the next. This map can be decomposed into spatial modes associated with characteristic temporal frequencies and growth/decay rates, offering both a predictive and interpretable model of the system's behavior.
In marine hydrodynamics, the application of DMD to the modeling of ship dynamics has been pioneered in \cite{diez2022datadriven}, in which the proof of concept of short-term forecasting of trajectories, motions, and loads of maneuvering ships in waves was given, without relying on governing equations.

While powerful, standard DMD assumes memory-less dynamics of a free dynamical system, and is thus limited in modeling history-dependent effects and systems influenced by external inputs.
To overcome these limitations, several extensions have been proposed. 
Hankel-DMD (HDMD) \cite{mezic2017} and Augmented-DMD \cite{serani2023} involve the augmentation of the system state with its time-lagged (delayed) copies and/or its derivatives to address the memory effects. 
The application of Augmented-DMD to the modeling of ship course keeping in irregular waves and performing turning circle in regular waves has been first addressed in \cite{serani2023}, showing improved results compared to standard DMD for ship motions and forces forecasting.
A similar framework named high-order DMD (HODMD) was used in \cite{chen2023,chen2024} for similar purposes. 
The use of HDMD for short-term forecasting of ship motions is statistically addressed in \cite{palma2024forecasting}, highlighting its potential for real-time prediction and control applications, and digital twinning.
Dynamic Mode Decomposition with Control (DMDc) is a methodological extension to DMD that incorporates control variables and forcing inputs in the system regression, which is particularly relevant for marine applications where control actions, such as rudder movements, and forcing inputs such as wave elevation, significantly impact the vessel's dynamics.
The first exploration into the use of DMDc as a data-driven system identification method to derive an input/output reduced order model for ship motions was conducted in \cite{Serani2024snh}
Input inclusion and Hankel extension have been combined in Hankel-DMD with control (HDMDc) and applied for the first time to ship motion and forces prediction in \cite{palma2025si}, which demonstrated the capability of the method to achieve good accuracies without degradation through the observation time.

A few approaches for introducing uncertainty quantification in DMD analysis have been presented in the literature so far. To this end, a probabilistic model was introduced in \cite{takeishi2017b} to deal with uncertainty in the measured data, assuming the presence of Gaussian noise in the snapshots (the probabilistic model corresponds to standard DMD in case of no noise in the data). The dynamic modes and their coefficients in the expansion are considered random variables with Gaussian prior, the posterior of dynamic modes and eigenvalues are then inferred using Gibbs sampling, providing uncertainty quantification.
Later, the bagging optimized-DMD was presented in \cite{Sashidhar2022}. Exploiting the optimized-DMD framework \cite{askham2018}, which uses a variable projection method for nonlinear least squares to compute the DMD for unevenly timed samples, the authors leverage Breiman’s statistical bagging sampling strategy over the available data snapshots to produce an ensemble of models and then evaluate uncertainty metrics for modes and eigenvalues.
A different approach has been recently used in \cite{palma2025windturbine} to introduce 
uncertainty quantification in HDMD and HDMDc predictions. 
Bayesian extensions of the two methods have been obtained by considering their hyperparameters as stochastic variables with a given prior, obtaining a posterior on the prediction through a Monte Carlo sampling. 
A Bayesian HDMD is proposed in \cite{palma2024forecasting} for real-time ship motion prediction, and \cite{palma2025si} extended the approach to the HDMDc, demonstrating improved accuracy and reliability in forecasting ship responses of the 5415M hull under severe sea near roll-resonance conditions.

Despite the advancements in both physics-based and data-driven modeling approaches, the need for accurate, interpretable, and computationally efficient models for seakeeping prediction remains pressing, particularly for complex platforms like catamarans operating in dynamic and uncertain environments.
Although DMD methods provide a compact and interpretable representation of the system’s dynamics and have been demonstrated to be well-suited for forecasting and system identification tasks, the literature review does not show contributions for multiple hull configurations. 
In addition, challenges remain in bringing prediction capabilities and uncertainty quantification together in a consistent, robust framework, especially when working with experimental data subject to measurement noise.

This paper aims to address this gap by proposing a novel ensemble HDMDc framework for predicting key seakeeping variables such as heave, pitch, vertical velocities and accelerations, and total resistance, based on irregular wave seakeeping basin experiments of the Delft 371 catamaran \cite{DURANTE2020} and including uncertainty estimation.
Wave elevation measured at a position longitudinally corresponding with the center of gravity (LCG) is used as the system forcing input. 
A regularized DMD regression introduced in \cite{XIE2024} is applied to deal with experimental data and measure noise, and obtain more accurate models.
The ensemble method is applied to introduce uncertainty quantification into the analysis, either with a Bayesian \cite{palma2024forecasting, palma2025si} (BHDMDc) or a Frequentist (FHDMDc) approach in a unified formulation, and the results of the two are compared. 
In the Bayesian approach, different models are obtained by varying the values of HDMDc hyperparameters: given their prior distributions, the posterior is evaluated on the prediction using Monte Carlo sampling. 
In the Frequentist approach, different models are obtained using several different subsamples of the training data. A mean prediction and its standard deviation are then calculated from the obtained ensemble. 
Quantifying the uncertainty in predictions, the ensemble HDMDc method provides both point forecasts and confidence intervals, which are critical for decision support in design and operations.

The predictions of the methods are tested to assess their time-resolved accuracy by evaluating the normalized root mean square error (NRMSE) between the prediction and reference EFD data for a set of test time histories. 
In addition, the statistics of the irregular wave time series are addressed by providing variables' probability density functions (PDF) from EFD, ensemble HDMDc, and CFD from literature \cite{diez2018statistical}. These are obtained through a moving block bootstrap method (MBB) and evaluating their differences using the Jensen-Shannon divergence (JSD) metric.

The paper is organized as follows; 
\cref{s:testcase} outlines the test case, \textit{i.e.}, the Delft 372 hull and the scaled replica for basin tests named CNR-INM 2554 and its setup; 
the formulations of HDMDc and its ensemble UQ extensions (namely the BHDMDc and the FHDMDc), the MBB, and the error metrics are detailed in \cref{s:methods};
\cref{s:numerical} reports the numerical setup for the DMD-based methods;
the results of the HDMDc, BHDMDc, and FHDMDc models are shown in \cref{s:results} and discussed in \cref{s:discussion}; finally, some concluding remarks can be found in \cref{s:conclusion}.

\section{Test Case}\label{s:testcase}
For this study, experimental data from seakeeping basin tests on the TU Delft 
372 catamaran \cite{vantveer1998} in irregular waves are used. 
The experiments have been conducted at the CNR-INM \textit{Emilio Castagneto} towing tank, which dimensions are 220 m, 9 m, and 3.6 m in length, width, and depth, respectively, and fully reported in \cite{DURANTE2020}. 
The tank is equipped with a towing carriage capable of operating at a maximum speed of 15.0 m/s. The head wave system is generated by means of a Kempf \& Remmers wave maker. The wave maker is equipped with a flap plunger hinged 1.80 m below the calm water level, with a maximum angle range $\pm13^{\circ}$, frequency range 0.1–1.8 Hz.
\begin{figure}[ht!]
\begin{center}
    \includegraphics[width=0.5\linewidth]{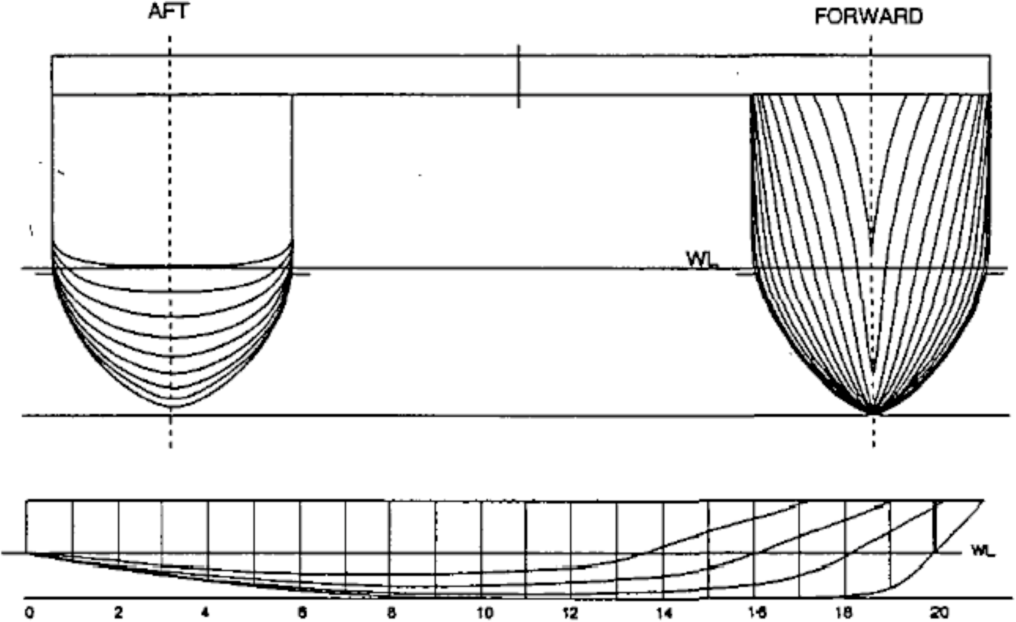}
    \caption{Delft 372 catamaran model: geometry and body plan.}
    \label{fig:hull}
\end{center}
\end{figure}
A fiberglass scale model, designated as CNR-INM 2554, was constructed to replicate the Delft design at a scale of 1:33.33, corresponding to a model length ($L_{pp}$) of 3 meters, considering the full-scale prototype to be a notional 100-meter catamaran. 
An outline of the model’s geometry and body plan is illustrated in \cref{fig:hull}, with a visual of the manufactured model shown in \cref{fig:model}. Key geometric, hydrostatic, and hydrodynamic parameters at both model and extrapolated full scales are summarized in \cref{tab:cnr_inm_model_2554}.

\begin{figure}[ht!]
\begin{center}
    \includegraphics[width=0.5\linewidth]{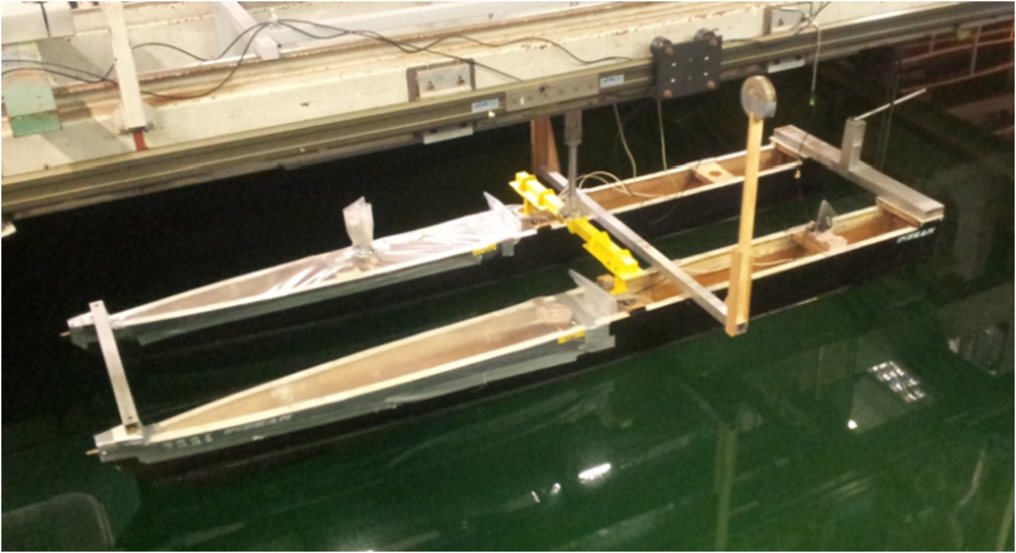}
    \caption{CNR-INM 2554 model for towing-tank seakeeping experiments.}
    \label{fig:model}
\end{center}
\end{figure}

The structural stiffness of the catamaran model was enhanced by rigidly connecting the demi-hulls with three transverse aluminum beams, two positioned fore and aft of the center of gravity, and a third one located at the aft perpendicular (see \cref{fig:model}). 
The rigidity of the demi-hulls arrangement did not require a connecting deck in the forward section, and hence prevented the occurrence of deck-slamming events during testing.
%
\begin{table}
\begin{center}
    \caption{Main particulars of the CNR-INM model 2554.}  
    \label{tab:cnr_inm_model_2554}      
\scalebox{0.68}{    
    \begin{tabular}{lllll}  
        \toprule  
        \textbf{Particular} & \textbf{Symbol} & \textbf{Unit} & \multicolumn{2}{l}{\textbf{Value}} \\  
        \cmidrule(l{2pt}r{2pt}){4-5}      
        & & & \textbf{Model} & \textbf{Full} \\  
        \midrule  
        Scale & $\Lambda$ & - & 33.33 & -\\  
        Length between perpendiculars & L$_{\text{pp}}$ & m & 3.00 & 100.00 \\  
        Beam overall & B & m & 0.94 & 31.33 \\  
        Beam demi-hull & b & m & 0.24 & 8.00 \\  
        Distance between center of the demi-hulls & H & m & 0.70 & 23.33 \\  
        Draught & T & m & 0.15 & 5.00 \\  
        Displacement & $\Delta$ & kg & 87.0 & 3225000 \\  
        Vertical center of gravity & KG & m & 0.34 & 11.33 \\  
        Longitudinal center of gravity & LG & m & 1.41 & 47.00 \\  
        Pitch radius of gyration & K$_\text{yy}$ & m & 0.782 & 26.07 \\  
        \bottomrule  
    \end{tabular}  
}
\end{center}
\end{table} 

The model was towed through a twin-gimbal, one for each demi-hull, rigidly connected with a transversal beam which ensured free pitch motion around the vessel's center of gravity. A vertical beam sliding in a linear bearing allowed free heave motion.
Other degrees of freedom such as surge, sway, and roll were mechanically constrained by the linking system, while the yaw angle is fixed by a motion restrictor.
However, the linking system was not perfectly rigid, resulting in elastic deformations of the mounting system. These deflections introduced higher-order vibration and noise in the measured resistance, particularly at elevated test speeds.

The North Pacific Ocean environment has been selected as the operative scenario to be replicated in the seakeeping test, which main characteristics are reported in \cref{tab:sea_states}: the mean wave height (trough to crest) of the highest third of the waves ($H_{1/3}$) and the most probable modal wave period ($T_m$) are shown for sea states from 1 to 8. 
The irregular wave system has been generated by considering the Bretschneider energy spectrum, which is the best representation of the present scenario, being developed for modeling the North Pacific area.

Following \cite{diez2013reliability}, the 100 m full-scale length catamaran (advancing in head waves) is considered to have a variable operational speed within the range from Fr$ = 0.115$ to $0.575$; the experiments were conducted considering an average Froude number of Fr = 0.425 and a sea state 5.
The main sea characteristics are indicated in \cref{tab:irregular_wave_parameters}: it is worth noticing that the maximum steepness is close to the linearity limit. 

The measured variables that are relevant to this study are the wave elevation at the LCG position $\eta_{cg}$, the free motion variables of the vessel, \textit{i.e.,} the heave $z$ and pitch $\theta$ of the model, the vertical velocity at a notional flight-deck $\dot z_D$, the vertical acceleration at the bridge $\ddot z_B$, and the total resistance in waves $F_x$.
The motion variables, particularly the location of the velocity and acceleration measures, have been selected as per the NATO standardization agreement STANAG 4154 \cite{Kennell1985, STANAG4154}.
\begin{table*}[htbp]
\begin{center}
    \caption{Characteristic annual values for sea states in the North Pacific.}
    \label{tab:sea_states}
\scalebox{0.85}{      
    \begin{tabular}{lllllllll}
        \toprule
        Sea State & Mean $H_{1/3}$ (m) & \multicolumn{4}{l}{Probability of sea state (\%)} & \multicolumn{3}{c}{Most probable modal wave period $T_m$ (s)} \\
        \cmidrule(lr){3-6} \cmidrule(lr){7-9}
        & & Bales, 1982 & Lee and Bales, 1984 & Average & Exceedance & Bales, 1982 & Lee and Bales, 1984 & Average\\
        \midrule
        0-1 & 0.05 & 0.00 & 1.30 & 0.65 & 99.35 & - & - & -\\
        2 & 0.30 & 4.10 & 6.40 & 5.25 & 94.10 & 7.50 & 6.30 & 6.90\\
        3 & 0.88 & 16.90 & 15.50 & 16.20 & 77.90 & 7.50 & 7.50 & 7.50\\
        4 & 1.88 & 27.80 & 31.60 & 29.70 & 48.20 & 8.80 & 8.80 & 8.80\\
        \textbf{5} & \textbf{3.25} & \textbf{23.50} & \textbf{20.94} & \textbf{22.22} & \textbf{25.98} & \textbf{9.70} & \textbf{9.70} & \textbf{9.70}\\
        6 & 5.00 & 16.30 & 15.03 & 15.67 & 10.32 & 13.80 & 12.40 & 13.10\\
        7 & 7.50 & 9.10 & 7.60 & 8.35 & 1.96 & 13.80 & 15.00 & 14.40\\
        8 & 11.50 & 2.20 & 1.56 & 1.88 & 0.08 & 18.00 & 16.40 & 17.20\\
        $>8$ & $>14$ & 0.10 & 0.07 & 0.09 & 0.00 & 20.00 & 20.00 & 20.00\\
        \bottomrule
    \end{tabular}
}
    \end{center}
\end{table*}
\begin{table}
\begin{center}    
    \caption{Irregular wave parameters.}
    \label{tab:irregular_wave_parameters}
\scalebox{0.7}{
    \begin{tabular}{llll}
        \toprule
        Quantity & Symbol & Full sc. & Model sc. \\
        \midrule
        Speed of Advancement & $U$ [m/s] & 13.3091 & 2.305 \\
        & $U$ [kn] & 25.8711 & 4.481 \\
        Froude Number & $Fr[-]$ & 0.425 & 0.425 \\
        Mean value of 1/3 largest wave & $H_{1/3}$ [m] & 3.2500 & 0.0975 \\
        Mean wave height & $H_m$ [m] & 2.0366 & 0.0611 \\
        RMS of wave height & $H_{rms}$ [m] & 2.2981 & 0.0689 \\
        Encountering frequency (min) & $f_e$ [Hz] min & 0.0800 & 0.4619 \\
        Encountering frequency (max) & $f_e$ [Hz] max & 0.4000 & 2.3094 \\
        Wave frequency (min) & $f$ [Hz] min & 0.0546 & 0.3077 \\
        Wave frequency (max) & $f$ [Hz] max & 0.1657 & 0.9596 \\
        Wave length (min) & $\lambda_w$ [m] min & 56.8146 & 1.7547 \\
        Wave length (max) & $\lambda_w$ [m] max & 523.7561 & 14.9574 \\
        Non dimensional wave length (min) & $\lambda_w/L_{pp}$ min & 0.5681 & 0.5682 \\
        Non dimensional wave length (max) & $\lambda_w/L_{pp}$ max & 5.2376 & 4.9858 \\
        Min $H_m$ over wave length & $H_m/\lambda_w$ min & 0.0039 & 0.0041 \\
        Max $H_m$ over wave length & $H_m/\lambda_w$ max & 0.0358 & 0.0358 \\
        Steepness (min) & $kA$ min & 0.0122 & 0.0128 \\
        Steepness (max) & $kA$ max & 0.1126 & 0.1126 \\
        \bottomrule
    \end{tabular}
}
\end{center}
\end{table}
%

The wave elevation $\eta_{cg}$ has been measured by means of a Kenek 
probe placed approximately 3 m aside the hull at the LCG position (see Fig. 5); 
The probe is a non-intrusive instrumentation with an accuracy of 0.1 mm and a maximum range of measurement of 150 mm. This probe is suitable for low-speed seakeeping tests and moderate wave steepness, therefore, it has been considered accurate enough for the wave steepness 
range and Fr number at which the present tests are carried on.

The free motions of the model variables $z$ and $\theta$ are measured through a Krypton optical system with an acquisition frequency of 800 Hz.
It consists of three linear CCD cameras (K600 camera unit), which detect the position of a reference system fixed to the body (identified through three infrared LEDs mounted in the stern region of the left hull). 
The spatial position of each LED is found with a high resolution through a triangulation procedure. 
In addition, $\ddot z_B$ and $\ddot z_D$ are measured by two accelerometers placed at the following positions: 
\begin{itemize}
    \item Bridge Accelerometer: Horizontal 0.30L$_\text{pp}$ (90 cm) from the Front Perpendicular (FP), Vertical 0.15L$_\text{pp}$ (45 cm) from the keel line
    \item Flight-Deck Accelerometer: Horizontal 0.85L$_\text{pp}$ (255 cm) from the FP, Vertical 0.10L$_\text{pp}$ (30 cm) from the keel line.
\end{itemize}
The notional flight-deck velocity $\dot z_D$ is obtained through time integration. 

The total resistance in waves $F_x$ is directly measured with two OMEGA LC204 load cells with a maximum range of 445N and accuracy 
around 0.1\%. 
The cells are placed between the gimbal and the ship model, one for each gimbal. 
Those are designed to be surface mounted with the load applied through the mounting stud. 

Additional details on the experiments and the setup can be found in \cite{DURANTE2020}.

\section{Methods}\label{s:methods}
\subsection{HDMDc}
DMD \cite{schmid2008dynamic,schmid2010} was originally presented to decompose high-dimensional time-resolved data into a set of spatiotemporal coherent structures, characterized by fixed spatial structures (modes) and associated temporal dynamics, providing a linear reduced-order representation of possibly nonlinear system dynamics.
The DMD operates equally on measured or simulated data and obtains its model with a direct procedure, constituting the training phase from a machine learning perspective, without requiring any specific knowledge or assumption about the system dynamics.

The data-driven nature, non-iterative training process, and data efficiency of DMD have contributed to its widespread application as a reduced-order modeling technique 
and real-time forecasting tool in various fields, including fluid dynamics, neuroscience, and structural health monitoring \cite{rowley2009, schmid2010, Tang2012, Semeraro2012, Song2013, Proctor2015, brunton2016, mann2016}. 
In naval contexts, DMD has been employed to analyze and extract knowledge from wave-induced motions \cite{diez2022datadriven, serani2023}, short-term forecasting of future vessel states in coursekeeping and maneuvering \cite{diez2022datadriven,palma2024forecasting}, and for system identification of coursekeeping ships \cite{palma2025si}.

The original DMD characterizes naturally evolving dynamical systems, while its extension to forced systems, called DMD with control (DMDc) \cite{proctor2016dynamic}, includes in the analysis the influence of forcing inputs and disambiguates it from the unforced dynamics of the system. 
The potential of the DMD in the analysis of nonlinear systems comes from its close relation to the Koopman operator \cite{Koopman1931,rowley2009}, 
which defines the possibility of transforming a nonlinear dynamical system into a possibly infinite-dimensional linear system \cite{Tu2014}. 
In this perspective, DMD is an equation-free data-driven approach that was shown in \cite{rowley2009} to be a computation of the Koopman operator for linear observables \cite{Marusic2024}.

The standard DMD and DMDc formulations approximate the Koopman operator based on linear measurements, creating a best-fit linear model linking sequential data snapshots \cite{schmid2010,kutz2016dynamic,mezic2022},
providing a locally linear representation of the dynamics that is unable to capture many essential features of nonlinear systems.
The augmentation of the system state is thus the subject of several DMD algorithmic variants 
aiming to find a coordinate system (or \textit{embedding}) that spans a Koopman-invariant subspace, to search for an approximation of the Koopman operator valid also far from fixed points and periodic orbits in a larger space.
However, there is no general rule for defining these observables and guaranteeing they will form a closed subspace under the Koopman operator \cite{brunton2016b}.

The HDMD \cite{mezic2017} is a specific version of the DMD algorithm developed to deal with the cases of nonlinear systems in which only partial observations are available \cite{Brunton2021}.
Incorporating time-lagged information in the data used to learn the model, HDMD and HDMDc increase the dimensionality of the system. 
Including time-delayed data in the analysis, the HDMD and HDMDc can extract linear modes spanning a space of augmented dimensionality that are able to reflect the non-linearities in the time evolution of the original system through complex relations between present and past states.
The state vector is thus augmented, embedding $s$ time-delayed copies of the original variables. 
The HDMDc involves, in addition, the augmentation of the input vector with $z$ time-delayed copies of the original forcing inputs. 
The use of time-delayed copies as additional observables in the DMD has been connected to the Koopman operator as a universal linearizing basis \cite{Brunton2017}, yielding the true Koopman eigenfunctions and eigenvalues in the limit of infinite-time observations \cite{mezic2017}.

The HDMDc approximates a dynamical system with external forcing as:
\begin{equation}\label{eq:hdmdcdsys}
    \mathbf{\hat{x}}_{j+1} = \widehat{\mathbf{A}}\mathbf{\hat{x}}_j + \widehat{\mathbf{B}}\mathbf{\hat{u}}_j.
\end{equation}
The vectors $\mathbf{\hat{x}}_j$ and $\mathbf{\hat{u}}_j$ are called the extended state and input vectors snapshot at the time instant $j$.
The definition of $\mathbf{\hat{x}}_j$ and $\mathbf{\hat{u}}_j$ is obtained starting from the original state $\mathbf{x}_j$ and input $\mathbf{u}_j$ vectors snapshots:
\begin{alignat}{2}
    \mathbf{x}_j &= 
    \begin{bmatrix}
    x_{1} \\ x_{2} \\ \dots \\ x_{n}
    \end{bmatrix} \in \mathbb{R}^{n},
\quad
    \mathbf{u}_j &= 
    \begin{bmatrix}
    u_{1} \\ u_{2} \\ \dots \\ u_{q}
    \end{bmatrix} \in \mathbb{R}^{q},
\end{alignat}
which are augmented by embedding $s$ and $z$ time-lagged (delayed) copies of the original state and input variables, such that:
\begin{align}
    \hat{\mathbf{x}}_j &= 
    \begin{bmatrix}
    \mathbf{x}_j \\ \mathbf{x}_{j-1} \\ \dots \\ \mathbf{x}_{j-s}
    \end{bmatrix} \in \mathbb{R}^{n(s+1)},
\quad
    \hat{\mathbf{u}}_j &= 
    \begin{bmatrix}
    \mathbf{u}_j \\ \mathbf{u}_{j-1} \\ \dots \\ \mathbf{u}_{j-z}
    \end{bmatrix} \in \mathbb{R}^{q(z+1)}.
\end{align}
As a consequence, 
$\widehat{\mathbf{A}} \in \mathbb{R}^{n (s+1) \times n (s+1)}$ and $\widehat{\mathbf{B}} \in \mathbb{R}^{n (s+1) \times q (z+1)}$ are the extended state matrix and the extended system input matrix, respectively.

Introducing the vector $\mathbf{\hat{y}}_j \in \mathbb{R}^{n(s+1)+q(z+1)}$
\begin{equation}\label{eq:Y}
\mathbf{\hat{y}}_j=
\begin{bmatrix}
\mathbf{\hat{x}}_j \\
\mathbf{\hat{u}}_j\\
\end{bmatrix},
\end{equation}
\cref{eq:hdmdcdsys} can be rewritten as:
\begin{equation}\label{eq:dmdSIY}
\mathbf{\hat{x}}_{j+1}=\mathbf{\widehat{G}\hat{y}}_j, \hspace{1cm} \text{with} \hspace{1cm} \mathbf{\widehat{G}}=
\begin{bmatrix}
    \mathbf{\widehat{A}} & \mathbf{\widehat{B}} \\
\end{bmatrix}.
\end{equation}
Data are collected for model building from $m$ snapshots and are arranged in two augmented data matrices $\widehat{\mathbf{Y}} \in \mathbb{R}^{(n(s+1)+q(z+1))\times (m-1)}$ and $\widehat{\mathbf{X}}' \in \mathbb{R}^{n(s+1)\times(m-1)}$, which are built as:
\begin{equation}\label{eq:scXX'}
\widehat{\mathbf{Y}}=
\begin{bmatrix}
\mathbf{X} \\
\mathbf{S}\\
\mathbf{U} \\
\mathbf{Z}\\
\end{bmatrix},
\qquad
\widehat{\mathbf{X}}'=
\begin{bmatrix}
\mathbf{X}' \\ 
\mathbf{S}'\\
\end{bmatrix}.
\end{equation}
The matrices $\mathbf{X} \in \mathbb{R}^{n \times (m-1)}$, $\mathbf{X}' \in \mathbb{R}^{n\times (m-1)}$, and $\mathbf{U} \in \mathbb{R}^{q \times (m-1)}$ contain the extended state and input snapshots at the $m$ considered time instants:
\begin{equation}\label{eq:XX'}
\begin{split}
\mathbf{X}=
\begin{bmatrix}
\mathbf{x}_j & \mathbf{x}_{j+1} & \dots & \mathbf{x}_{m-1}\\
\end{bmatrix},\\
\mathbf{X}'=
\begin{bmatrix}
\mathbf{x}_{j+1} & \mathbf{x}_{j+2} & \dots & \mathbf{x}_{m}\\
\end{bmatrix},\\
\mathbf{U}=
\begin{bmatrix}
\mathbf{u}_j & \mathbf{u}_{j+1} & \dots & \mathbf{u}_{m-1}\\
\end{bmatrix},
\end{split}
\end{equation}
while the Hankel matrices $\mathbf{S}$, $\mathbf{S}'$, and $\mathbf{Z}$ contain the lagged (delayed) extended state and input snapshots:  
\begin{align}\label{eq:s}
\mathbf{S}&=
\begin{bmatrix}
\mathbf{x}_{j-1} & \mathbf{x}_{j} & \dots & \mathbf{x}_{m-2}\\
\mathbf{x}_{j-2} & \mathbf{x}_{j-1} & \dots & \mathbf{x}_{m-3}\\
\vdots & \vdots & \vdots & \vdots \\
\mathbf{x}_{j-s-1} & \mathbf{x}_{j-s} & \dots & \mathbf{x}_{m-s-1}
\end{bmatrix}, 
\end{align}
\begin{align}\label{eq:s'}
\mathbf{S}'&=
\begin{bmatrix}
\mathbf{x}_{j} & \mathbf{x}_{j+1} & \dots & \mathbf{x}_{m-1}\\
\mathbf{x}_{j-1} & \mathbf{x}_{j} & \dots & \mathbf{x}_{m-2}\\
\vdots & \vdots & \vdots & \vdots \\
\mathbf{x}_{j-s} & \mathbf{x}_{j-s+1} & \dots & \mathbf{x}_{m-s}
&\end{bmatrix},
\end{align}
\begin{align}\label{eq:z}
\mathbf{Z}&=
\begin{bmatrix}
\mathbf{u}_{j-1} & \mathbf{u}_{j} & \dots & \mathbf{u}_{m-2}\\
\mathbf{u}_{j-2} & \mathbf{u}_{j-1} & \dots & \mathbf{u}_{m-3}\\
\vdots & \vdots & \vdots & \vdots \\
\mathbf{u}_{j-z-1} & \mathbf{u}_{j-z} & \dots & \mathbf{u}_{m-z-1}\\
\end{bmatrix}.    
\end{align}
The augmented matrix $\widehat{\mathbf{G}} = [\widehat{\mathbf{A}}\; \widehat{\mathbf{B}}] \in \mathbb{R}^{n(s+1) \times(n(s+1)+q(z+1))}$ is approximated by HDMDc by solving the following regularized least-square minimization \cite{XIE2024}:
\begin{equation}
    \min_{\widehat{A},\widehat{B}} \quad || \widehat{\mathbf{X}}' - \mathbf{[\widehat{\mathbf{A}}\; \widehat{\mathbf{B}}]} \; \widehat{\mathbf{Y}} ||^2_{F} + \lambda  \mathbf{[\widehat{\mathbf{A}}\; \widehat{\mathbf{B}}]} ||^2_{F},
\end{equation}
where $\lambda$ is a regularization factor, which solution is given by:
\begin{equation}
   [\widehat{\mathbf{A}}\; \widehat{\mathbf{B}}] = \widehat{\mathbf{X}}' \widehat{\mathbf{Y}}^{\text{T}} \left[ \widehat{\mathbf{Y}}\widehat{\mathbf{Y}}^{\text{T}} + \lambda \mathbf{I} \right]^{-1}.
\end{equation}
The Tikhonov-regularized formulation is suitable for improving the numerical stability of the DMD regression, its robustness in high-dimensional spaces as per the Hankel extension, and improving accuracy for applications to noisy data. 
Once the matrices $\widehat{\mathbf{A}}$ and $\widehat{\mathbf{B}}$ are obtained, \cref{eq:hdmdcdsys} is used to calculate the time evolution of the augmented state vector and, finally, by isolating its first $n$ components, the predicted time evolution of the original state variables $\mathbf{{x}}(t)$ is extracted.

\subsection{Uncertainty quantification in HDMDc through ensembling}
DMD-based models can be further extended to provide some uncertainty estimation of their predictions.
To this aim, here we use the ensembling approach, \textit{i.e.,} the combination of predictions coming from different models to obtain a mean prediction and its statistics.
In particular, inside the ensembling framework, we identify two DMD extensions. 

The first is called Bayesian extension, it has been first introduced in \cite{palma2025windturbine}, and also adopted in \cite{palma2024forecasting} and \cite{palma2025si} for both HDMD and HDMDc, leading to BHDMD and BHDMDc, respectively.
As highlighted by the authors in previous works \cite{diez2022snh, serani2023, Diez2024, Serani2024snh}, the final prediction from DMD-based models may strongly vary for different hyperparameter settings, and no general rule is given for the determination of their optimal values. 
In BHDMDc, the hyperparameters of the deterministic method are considered as stochastic variables with a given prior, and a posterior on the prediction is obtained through Monte Carlo sampling. Details of the formulation are given in \cref{s:bhdmdc}.

The second is here introduced for the first time as the Frequentist HDMDc (FHDMDc), as opposed to the Bayesian approach. Indeed, in the FHDMDc there are no stochastic variables with a given prior. Instead, uncertainty is obtained by combining the predictions of different HDMDc models trained on various training datasets. The FHDMDc formulation is detailed in \cref{s:fhdmdc}

\subsubsection{Bayesian HDMDc}\label{s:bhdmdc}
The definition of BHDMDc starts by noting that the dimensions and the values within matrices $\widehat{\mathbf{A}} \in \mathbb{R}^{n(s+1) \times n(s+1)}$ and $\widehat{\mathbf{B}} \in \mathbb{R}^{n(s+1) \times q(z+1)}$ depends on four hyperparameters of the algorithms: the observation time length, $l_{tr} = t_m - t_1$ , the maximum delay time in the augmented state $l_{d_x} = t_{j-1} - t_{j-s-1}$, the maximum delay time in the augmented input $l_{d_u} = t_{j-1} - t_{j-z-1}$ (or the number of snapshots $m$, the number of time-lagged state and input copies $s$ and $z$ in the discrete domain), and the regularization coefficient $\lambda$.

These dependencies can be denoted as follows:
\begin{equation}\label{eq:bayes1}
\begin{split}
\quad \widehat{\mathbf{A}}&=\widehat{\mathbf{A}}(l_{tr},l_{d_x},l_{d_u},\lambda), \qquad \\
\widehat{\mathbf{B}}&=\widehat{\mathbf{B}}(l_{tr},l_{d_x},l_{d_u},\lambda).
\end{split}
\end{equation}
In the Bayesian formulations, the hyperparameters are considered stochastic variables with given probability density functions $p(l_{tr})$, $p(l_{d_x})$, $p(l_{d_u})$, and $p(\lambda)$, introducing uncertainty in the process.
Through uncertainty propagation, the solution $\mathbf{x}(t)$ also depends on $l_{tr}$, $l_{d_x}$, $l_{d_u}$ and $\lambda$.
At a given time $t$, the expected value of the solution and its standard deviation can be expressed as:
%
\begin{equation}\label{eq:bayes3c}
\boldsymbol{\mu_x}(t)=\int_{\lambda^l}^{\lambda^u} \int_{l_{d_u}^l}^{l_{d_u}^u} \int_{l_{d_x}^l}^{l_{d_x}^u} \int_{l_{tr}^l}^{l_{tr}^u}\mathbf{x}(t,l_{tr},l_{d_x},l_{d_u},\lambda)\\p(l_{tr})p(l_{d_x})p(l_{d_u})p(\lambda)dl_{tr} dl_{d_x} dl_{d_u} d\lambda,
\end{equation}
%
\begin{equation}\label{eq:bayes4c}
\boldsymbol{\sigma_x}(t)= \Bigg\{ \int_{\lambda^l}^{\lambda^u} \int_{l_{d_u}^l}^{l_{d_u}^u} \int_{l_{d_x}^l}^{l_{d_x}^u} 
 \int_{l_{tr}^l}^{l_{tr}^u} \left[\mathbf{x}(t,l_{tr},l_{d_x},l_{d_u},\lambda)-\boldsymbol{\mu_x}(t) \right]^2 \\ p(l_{tr})p(l_{d_x})p(l_{d_u})p(\lambda)dl_{tr} dl_{d_x} dl_{d_u} d\lambda\Bigg\}^\frac{1}{2},
\end{equation}
%
where ${l_{tr}^l}$, ${l_{d_x}^l}$, ${l_{d_u}^l}$, $\lambda^l$ and ${l_{tr}^u}$, ${l_{d_x}^u}$, ${l_{d_u}^u}$, $\lambda^u$ are lower and upper bounds for $l_{tr}$, $l_{d_x}$, $l_{d_u}$, and $\lambda$.

In practice, a uniform probability density function is assigned to the hyperparameters, and a set of realizations is obtained through a Monte Carlo sampling. Accordingly, for a given test time sequence, the solution $\mathbf{x}(t,l_{tr},l_{d_x},l_{d_u},\lambda)$ is computed for each realization of the hyperparameters, and at a given time $t$ the expected value and standard deviation of the solution are then evaluated.

\subsubsection{Frequentist HDMDc}\label{s:fhdmdc}
In FHDMDc multiple observation datasets are defined. For each of those, a different HDMDc system with its own matrices $\widehat{\textbf{A}}(l_{tr},l_{d_x},l_{d_u},\lambda)$ and $\widehat{\textbf{B}}(l_{tr},l_{d_x},l_{d_u},\lambda)$ are obtained for a single hyperparameter set previously identified.
For any given test time sequence, hence, multiple solutions $\mathbf{x}_{i}(t,l_{tr},l_{d_x},l_{d_u},\lambda)$ are obtained, where $i\in[1,n_f]$ and $n_f$ indicates the different models available, one per each observation dataset.
The sample mean of the predictions can be calculated from the ensemble, along with the sample standard deviation:
\begin{equation}
    \boldsymbol{\mu_x}(t) = \frac{1}{n_f}\sum_{i=1}^{n_f} \mathbf{x}_{i}(t,l_{tr},l_{d_x},l_{d_u},\lambda)
\end{equation}
\begin{equation}
    \boldsymbol{\sigma_x}(t) = \left\{ \frac{1}{n_f-1} \sum_{i=1}^{n_f} \left[ \mathbf{x}_{i}(t,l_{tr},l_{d_x},l_{d_u},\lambda) - \boldsymbol{\mu_x}(t) \right] \right\}^{\frac{1}{2}}
\end{equation}

\subsection{Performance metrics}
To evaluate the time-resolved predictions from HDMDc, BHDMDc, and FHDMDc the normalized mean square error (NRMSE) \cite{Diez2024} is employed.
The NRMSE quantifies the average root mean square error between the predicted values $\mathrm{\mathbf{\tilde x}}_t$ and the measured (test) values $\mathrm{\mathbf{x}}_t$ at different time steps. 
It is calculated by taking the square root of the average squared differences, normalized by $k$ times the standard deviation of the measured values:
\begin{equation}\label{eq:nrmse}
   \mathrm{NRMSE} = \frac{1}{N} \sum_{i=1}^{N} \frac{1}{k\sigma_{x_i}}\sqrt{\frac{1}{\mathcal{T} } \sum_{j=1}^{\mathcal{T}} \left( \tilde{x}_{ij} - x_{ij} \right)^2},
\end{equation}
where $N$ is the number of variables in the predicted state, $\mathcal{T}$ is the number of considered time instants, and $\sigma_{x_i}$ is the standard deviation of the measured values in the considered time window for the variable $x_i$.
The NRMSE evidences the cumulative error due to phase, frequency, and amplitude deviations between the reference and the predicted signal, evaluating a pointwise difference between the two. 

In addition to the direct comparison of DMD-predicted and experimentally measured time histories, the calculation of the probability density function (PDF) of the variables object of prediction is of interest.
To statistically assess the PDFs of the variables, obtaining an expected value and confidence intervals, a MBB method is applied to time histories from EFD, CFD, and DMD-based predictions. 
Starting from a time signal with $\mathcal{T}$ samples, 
a number $C = \mathcal{T}-l+1$ of moving blocks is used, each defining a time history with $\xi_i$ time samples $i=c,\dots,c+l-1$, where $c$ is the block index and $l = (2\varphi/a)^{2/3}\mathcal{T}^{1/3}$ is an optimal block length \cite{carlstein1986use} with
\begin{equation}
    \varphi = \dfrac{\mathcal{T}\displaystyle\sum_{i=1}^{\mathcal{T}} \left[ \xi_{i+1} - \text{EV}(\xi) \right] \left[ \xi_{i} - \text{EV}(\xi) \right] }{(\mathcal{T}-1)\displaystyle\sum_{i=1}^{\mathcal{T}} \left[\left( \xi_{i} - \text{EV}(\xi) \right) \right]}
\end{equation}
and $a=(1-\varphi)(1+\varphi)$. 
For the MBB analysis, a single long time history of $\mathcal{T}$ samples is obtained for the EFD the CFD, and the FHDMDc by joining all the test sequences, the simulations and the predicted time series.
From the original set of C blocks, a number of $C' = \mathcal{T}/l$ blocks are drawn at random with replacement and concatenated in the order they are picked, forming a new bootstrapped series of size $\mathcal{T}$. 

The PDF of each bootstrapped time series 
is obtained using kernel density estimation \cite{Miecznikowski2010} as follows:
\begin{equation}
    \text{PDF}\left(\xi,y\right) = \frac{1}{\mathcal{T} h}\sum_{i=1}^{\mathcal{T}} K \left(\frac{y -\xi_i}{h}\right).
\end{equation}
Here, $K$ is a normal kernel function defined as
\begin{equation}
    K\left(\xi\right) = \frac{1}{\sqrt{2 \pi}} \exp{\left(-\frac{\xi^2}{2}\right)},
\end{equation}
where $h=1.06 \min\left(\sigma_x,\text{IQR}(\xi)\right)\mathcal{T}^{-1/5}$ is the bandwidth \cite{Silverman2018}. 
A total of B=100 bootstrapped series are used here, and hence, a set of 100 PDFs is obtained for each variable of the system state $\mathbf{x}$. 
The expected value and a confidence interval are calculated for the PDF of each variable for the EFD measurements \cite{DURANTE2020}, CFD simulations \cite{diez2018statistical}, and DMD-based predictions.
The quantile function $q$ is evaluated at probabilities $p=0.025$ and $0.975$, defining the lower and upper bounds of the 95\% confidence interval of the PDFs as $U_{\text{PDF}(\xi,y)} = \text{PDF}(\xi,y)_{q=0.95} - \text{PDF}(\xi,y)_{q=0.025}$. 

The so-obtained PDFs from the different sources are then compared using the Jensen-Shannon divergence (JSD) \cite{marlantes2024}. 
The JSD measures the similarity between two probability distributions as it estimates the entropy of the first probability density function $Q$ relative to the probability density function of the second $R$, where $M$ is the average of the two \cite{Lin1991}. 
\begin{align}
    &\mathrm{JSD} = \frac{1}{N} \sum_{i=1}^{N} \left( \frac{1}{2}D(Q_i\,||\,M_i) + \frac{1}{2}D(R_i\,||\,M_i) \right),  \quad \\
    &\quad \text{with} \quad M = \frac{1}{2} (Q + R)\label{eq:jsd}, \\
    &\quad \text{and} \quad D(K\,||\,H)=\sum_{y \in \chi} K(y) \ln\left( \frac{K(y)}{H(y)} \right). \label{eq:kld}
\end{align}
The JSD is based on the Kullback-Leibler divergence $D$, given by \cref{eq:kld}, which is 
the expectation of the logarithmic difference between the probabilities $K$ and $H$, both defined over the domain $\chi$, where the expectation is taken using the probabilities $K$ \cite{Kullback1951}
The similarity between the distributions is higher when the Jensen-Shannon distance is closer to zero. JSD is upper bounded by $\ln(2)$.
When applied to the PDFs evaluated from the bootstrapped time series, the expected value and the quantile function of JSD for $p=0.025$ and $0.975$, defining the lower and upper bound of the 95\% confidence interval $U_{\text{JSD}(Q,R)} = \text{JSD}(Q,R)_{q=0.975} - \text{JSD}(Q,R)_{q=0.025}$, are evaluated.
%

\section{Numerical setup for HDMDc}\label{s:numerical}
The HDMDc, BHDMDc, and FHDMDc are applied to the prediction of the state variables $\mathbf{x}=\left[z,\,\theta,\,\dot{z}_D,\,\ddot{z}_B,\,F_x \right]^{\text{T}}$, using $\mathbf{u}=\eta_{cg}$ as input for the identified dynamical system of \cref{eq:hdmdcdsys}.

Experimental data were collected from $n_\text{EFD}= 18$ independent EFD runs, each one spanning about $\mathcal{T}_i=5000$ time steps over approximately 80 encountered waves.
An average encounter wave period $\hat{T}=0.62s$ is calculated as the average zero-up crossing period of the measured $\eta_{\text{bow}}$ signal in the EFD runs:
\begin{equation}
    \hat{T} = \frac{1}{n_\text{EFD}}\sum_{i=1}^{n_\text{EFD}}\frac{N{z_{up_i}}}{\mathcal{T}_i}.
\end{equation}
For processing ease with DMD, data were downsampled to 64 time steps per $\hat{T}$.

Nine EFD runs were reserved as training data for building the HDMDc-based models. Five runs are used as a validation set to assess the effect of the hyperparameters in the deterministic HDMDc, 
and the remaining 4 runs are available as a test set to evaluate the performance of HDMDc and its UQ extensions. 

A full factorial exploration in the HDMDc hyperparameters domain is performed, using 3 levels per parameter, namely $l_{tr}/\hat{T} = 5,\,10,\,20$, $l_{d_x}/\hat{T} = 0,\,1,\,2$, $l_{d_u}/\hat{T} = 0,\,1,\,2$, and $\lambda = 10,\,100,\,1000$.
One time sequence is extracted from each training EFD run, generating 9 HDMDc models per hyperparameters combination.
Two time sequences are extracted from each validation and test EFD run for evaluating NRMSE and for MBB analysis.
Each hyperparameter combination leads to a different HDMDc model in terms of $\widehat{\mathbf{A}}$ and $\widehat{\mathbf{B}}$, which is thus assessed (\textit{i.e.,} the NRMSE of its predictions is evaluated) over 81 combinations of training and validation signals.
All the validation and test time histories are taken of length $l_{te}=15\hat{T}$.

All the analyses are based on normalized data using the Z-score standardization. Specifically, the time histories of each variable are shifted and scaled using the average and standard deviation evaluated on the training signal.

The results of the deterministic analysis are used to set up the uncertainty-aware BHDMDc and FHDMDc. 
In particular, a suitable range of variation of the hyperparameter is selected for the BHDMDc by identifying combinations showing low average NRMSE over the validation time sequences. 
A uniform probability is assigned to the hyperparameters in their variation range and 100 Monte Carlo realizations, using a random training time series, are used to obtain the posterior for the prediction (the actual number of delays $n_{d_x}$ and $n_{d_u}$ and training samples $n_{tr}$ are taken as the nearest integers from the calculated values of $l_{tr},\,l_{d_x}$, and $l_{d_u}$). Each Monte Carlo sample identifies a different HDMDc model of the Bayesian ensemble, and their predicted time histories are combined in a mean prediction with uncertainty for each test sequence.

On the other hand, the FHDMDc model is defined by the ensemble of the 9 HDMDc models built with the different training signals and using the hyperparameters' combination characterized by the lowest average NRMSE value from the deterministic analysis over the validation time sequences.



\section{Results}\label{s:results}
\begin{figure}[htbp]
    \begin{center}
    \includegraphics[width=0.9\linewidth]{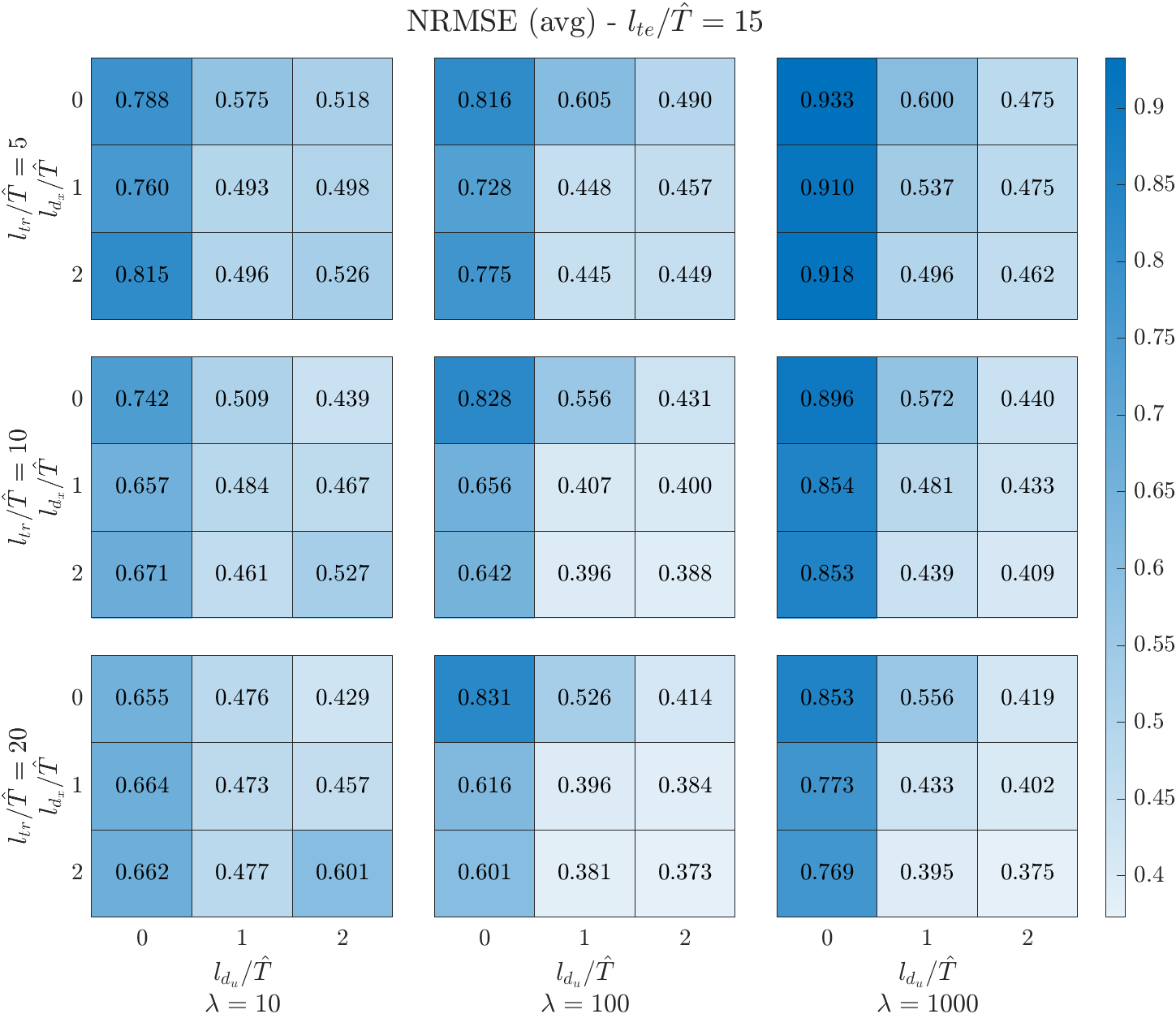}
    \caption{Average value of NRMSE for HDMDc over validation sequences varying hyperparameters' values, full factorial analysis.}
    \label{fig:hdmdc-nrmse-ave}
    \end{center}
\end{figure}

\begin{figure}[htbp]
    \begin{center}
    \includegraphics[width=0.9\linewidth]{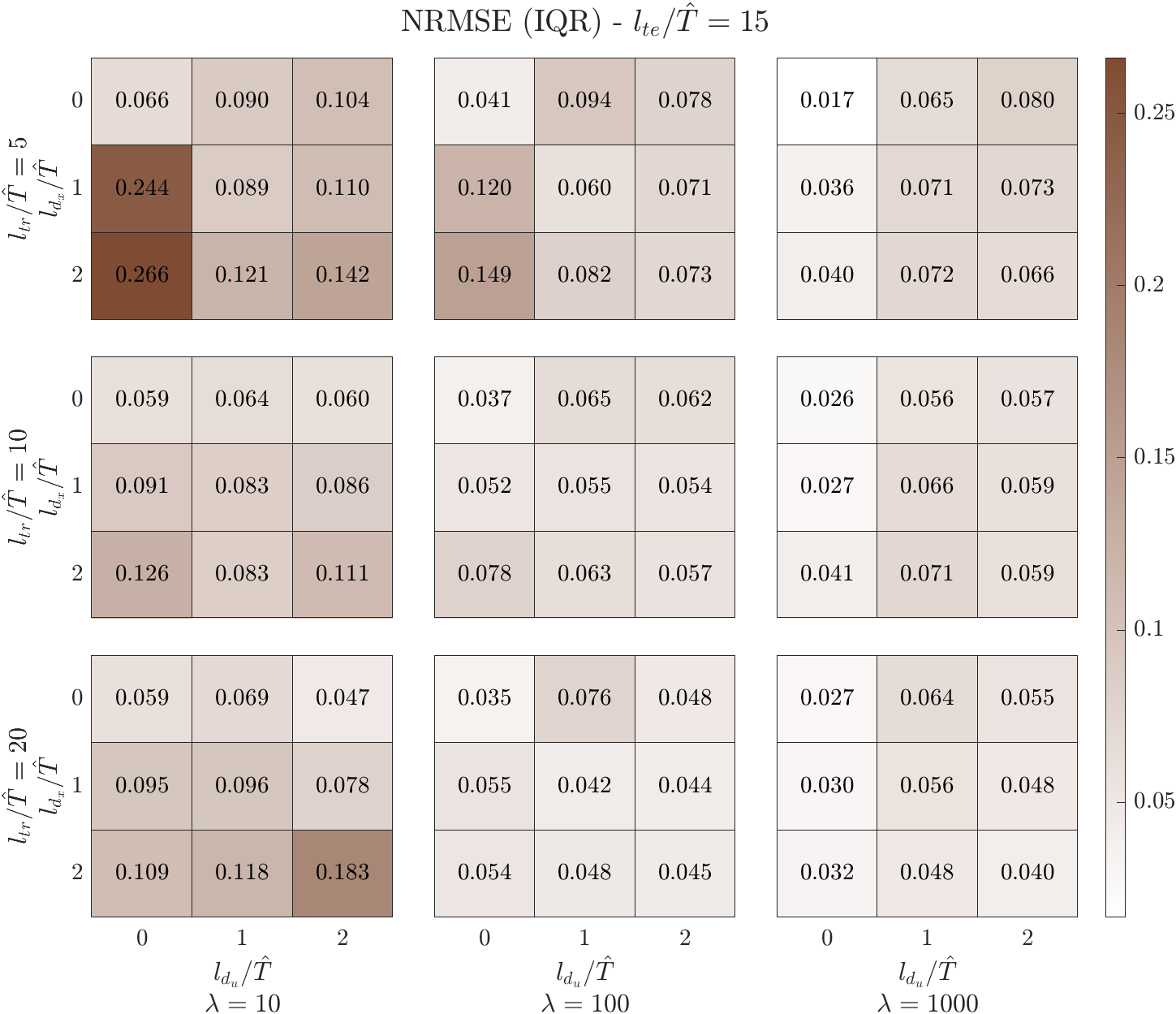}
    \caption{Interquartile range of NRMSE for HDMDc over validation sequences varying hyperparameters' values, full factorial analysis.}
    \label{fig:hdmdc-nrmse-iqr}
    \end{center}
\end{figure}

This section reports the results obtained with the HDMDc and its two uncertainty-aware extensions, BHDMDc and FHDMDc. 
The NRMSE scores obtained for the full factorial analysis with HDMDc varying the hyperparameters' values are summarized in the heatmaps of \cref{fig:hdmdc-nrmse-ave} and \cref{fig:hdmdc-nrmse-iqr} in terms of mean value and interquartile range (IQR, defined as the difference between the third and the first quartiles), respectively, evaluated over the validation time sequences.

From the analysis of \cref{fig:hdmdc-nrmse-ave,fig:hdmdc-nrmse-iqr}, the ranges $10\le l_{tr}/\hat{T}\le 20$, $1\le l_{d_x}/\hat{T}\le 2$,$1\le l_{d_u}/\hat{T}\le 2$, $100\le \lambda \le 1000$ emerge as combinations providing good accuracy with small dispersion. These ranges are used to define the prior on the hyperparameters for the BHDMDc.
The hyperparameters configuration $l_{tr}=20\hat{T}$, $l_{d_x}=2\hat{T}$, $l_{d_u}=2\hat{T}$, and $\lambda=100$, also indicated in the following as the \textit{best NRMSE configuration}, stands out for reaching the lowest value of the average NRMSE, and will be used in the FHDMDc. 

\begin{figure}[htbp]
    \begin{center}
    \includegraphics[width=0.9\linewidth]{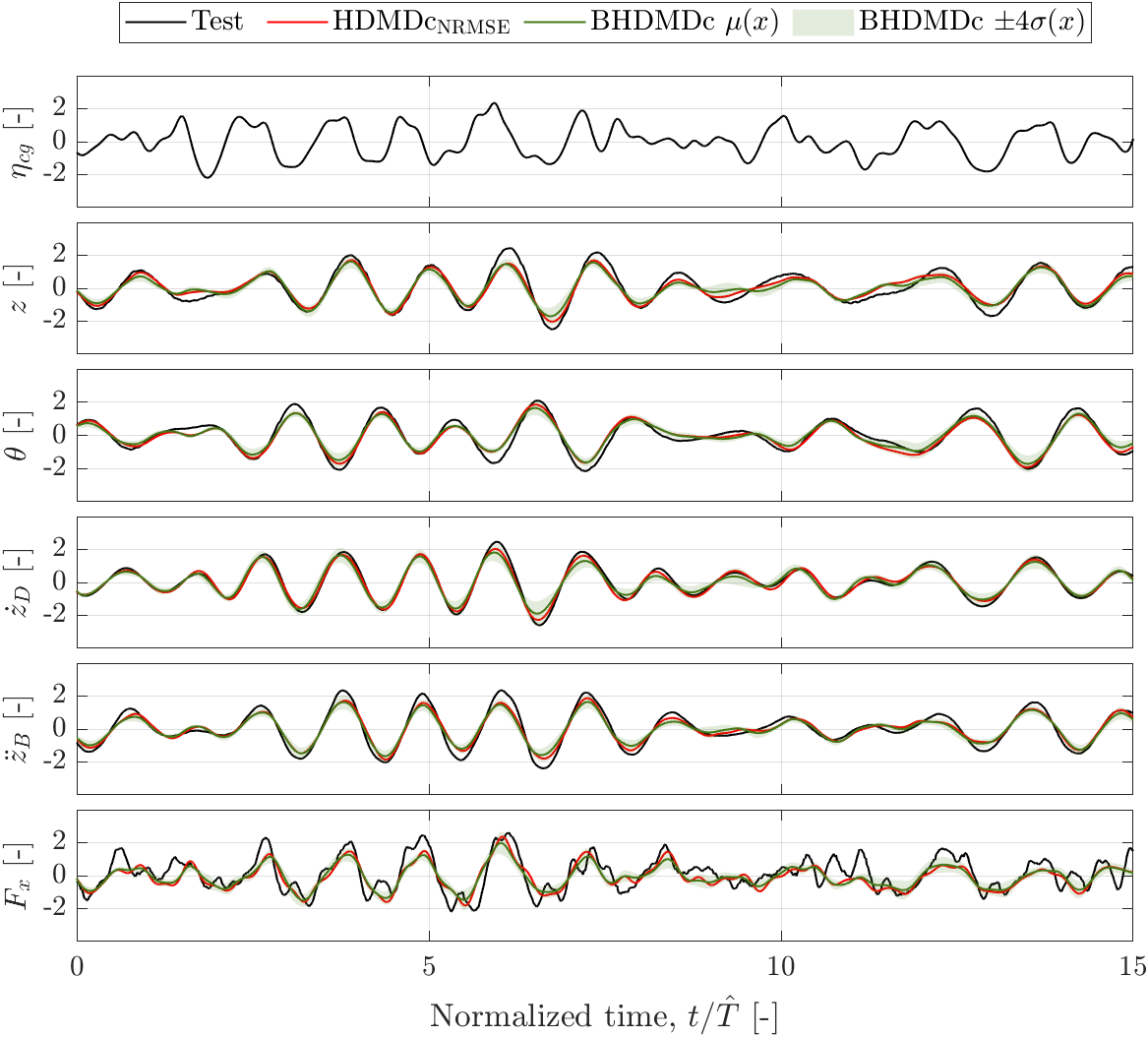}
    \caption{Time series prediction, EFD vs HDMDc vs BHDMDc, randomly selected test sequence.}
    \label{fig:BHDMDc_1}
    \end{center}
\end{figure}

\begin{figure}[htbp]
    \begin{center}
    \includegraphics[width=0.9\linewidth]{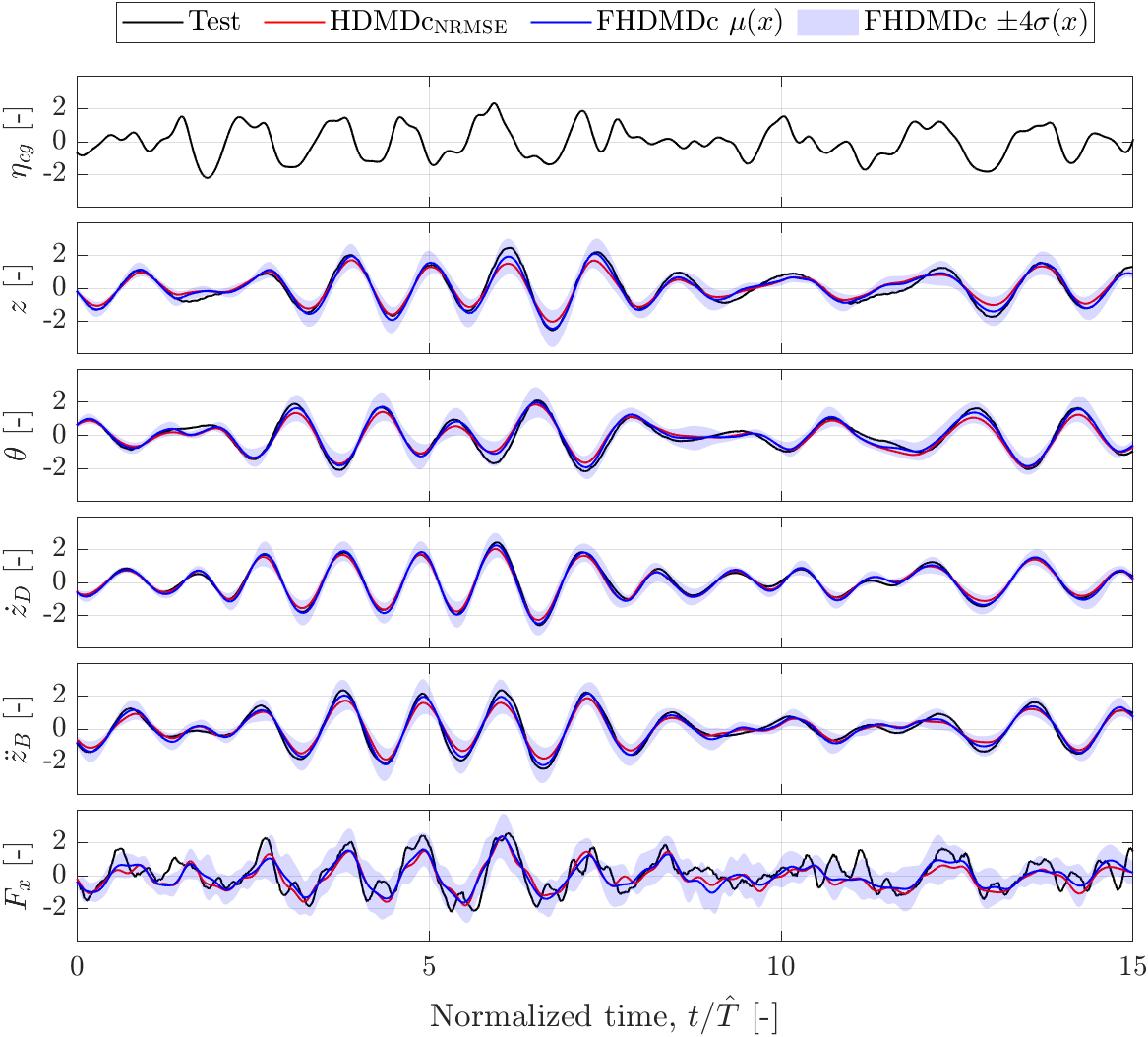}
    \caption{Time series prediction, EFD vs HDMDc vs FHDMDc, randomly selected test sequence.}    \label{fig:FHDMDc_1}
    \end{center}
\end{figure}

A direct visual comparison between the prediction obtained by HDMDc using the best NRMSE configuration  (HDMDc$_{\text{NRMSE}}$) and the EFD reference for a test sequence randomly selected from the test set is shown along with the prediction by BHDMDc and FHDMDc in \cref{fig:BHDMDc_1} and \cref{fig:FHDMDc_1}, respectively. Uncertainties for the ensemble methods are plotted as shaded areas covering the interval $\pm 4\sigma(\mathbf{x})$, while solid lines are used for the expected value of the predictions.

\begin{figure}[ht!]
    \begin{center}
    \includegraphics[width=0.5\linewidth]{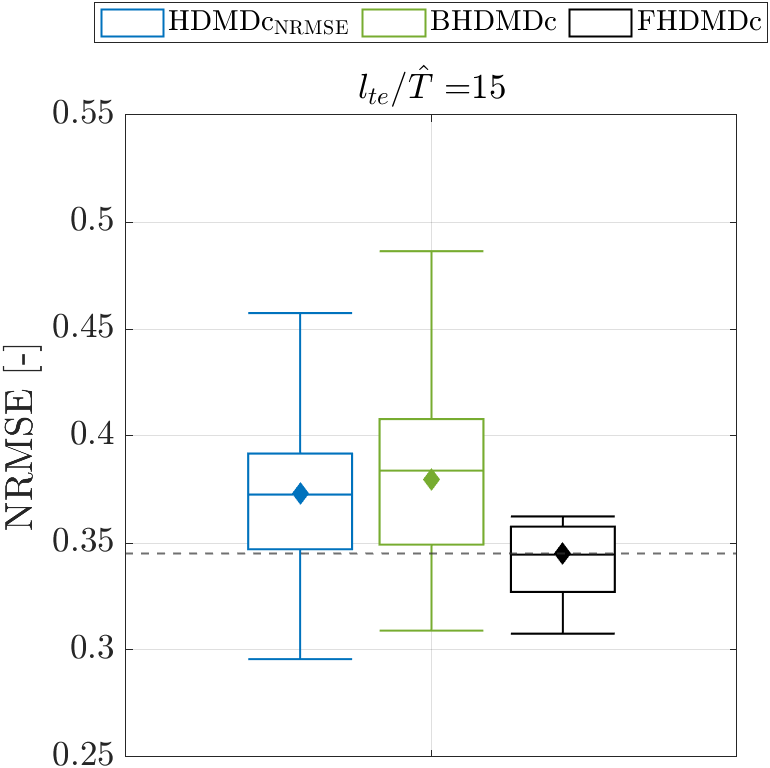}
    \caption{NRMSE box-and-whisker, comparison of HDMDc with best NRMSE hyperparameters, BHDMDc and FHDMDc, results over the test set.}
    \label{fig:detvsuq}
    \end{center}
\end{figure}

The accuracy of the HDMDc is compared with the BHDMDc and FHDMDc by evaluating the NRMSE on the test sequences. The results are shown in a box-and-whisker diagram in \cref{fig:detvsuq}. The boxes show the first, second (equivalent to the median value), and third quartiles, while the whiskers extend from the box to the farthest data point lying within 1.5 times the IQR. Diamonds indicate the average value of the results for each method. Outliers are not shown to improve the readability of the plot.

\begin{figure}[htbp]
    \begin{center}
        \includegraphics[width=0.9\linewidth]{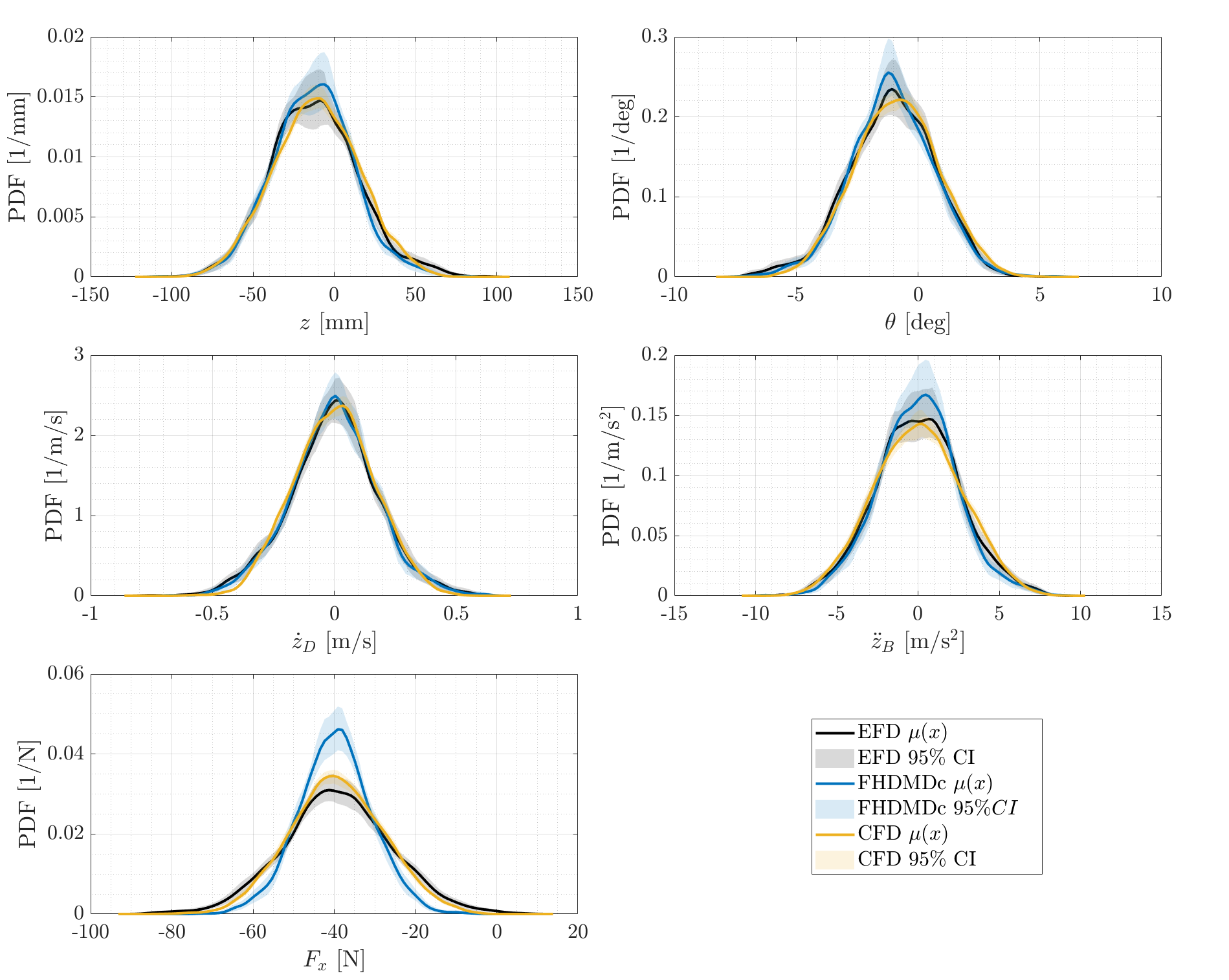}
        \caption{PDF estimation on bootstrapped time series for predicted variables. Comparison between EFD data, FHDMDc predictions, and CFD simulations from \cite{diez2018statistical}.}
        \label{fig:mbb}
    \end{center}
\end{figure}

Finally, the results from the MBB analysis applied to EFD and FHDMDc time series are shown in \cref{fig:mbb}. 
PDFs obtained from bootstrapped sequences are plotted for each predicted variable, along with their 95\% confidence interval.
CFD-based PDFs evaluated on simulation data available in literature \cite{diez2018statistical} are added to the plot for further comparison.
The differences between the distributions are assessed by evaluating the JSD between the EFD PDF and either the FHDMDc or CFD PDFs. 
The expected value and the quantile function $q$ applied to JSD evaluated at probabilities $p=0.025$ and $0.975$ are reported in \cref{tab:mbbjsd}, along with the uncertainty U.
\begin{table*}[htbp]
    \begin{center}
    \caption{Expected value and 95\% confidence lower bound, upper bound, and interval of JSD of EFD vs. FHDMDc PDFs and EFD vs. CFD PDFs evaluated on bootstrapped time series.}
    \label{tab:mbbjsd}    
    \begin{tabular}{lllll llll}
    \toprule
                & \multicolumn{4}{l}{JSD$_{\xi}$(EFD,FHDMDc)} & \multicolumn{4}{l}{JSD$_{\xi}$(EFD,CFD)}\\
    \cmidrule(l{2pt}r{12pt}){2-5} \cmidrule(l{2pt}r{12pt}){6-9}
    $\xi$       & EV  & q=0.025& q=0.975 & U & EV  & q=0.025& q=0.975 & U \\
             \midrule
    $z$         & 0.0053 & 0.0022 & 0.0126 & 0.0104 & 0.0042 & 0.0023 & 0.0068 & 0.0044 \\
    $\theta$    & 0.0039 & 0.0017 & 0.0097 & 0.0080 & 0.0056 & 0.0039 & 0.0087 & 0.0047 \\
    $\dot{z}_D$ & 0.0042 & 0.0013 & 0.0118 & 0.0106 & 0.0027 & 0.0008 & 0.0082 & 0.0074 \\
    $\ddot{z}_B$& 0.0031 & 0.0012 & 0.0073 & 0.0062 & 0.0060 & 0.0025 & 0.0112 & 0.0087 \\
    $F_x$       & 0.0346 & 0.0183 & 0.0541 & 0.0358 & 0.0073 & 0.0029 & 0.0133 & 0.0104 \\
    \midrule
    avg.        & 0.0085 & 0.0049 & 0.0159 & 0.0118 & 0.0052 & 0.0021 & 0.0080 & 0.0059 \\
    \midrule
    \end{tabular}
    \end{center}
\end{table*}

\section{Discussion}\label{s:discussion}
The analysis of \cref{fig:hdmdc-nrmse-ave,fig:hdmdc-nrmse-iqr} gives some hints about the role of the hyperparameter $\lambda$ in the regularization of the DMD-based regression and its tuning.
A reduction in the IQR amplitude of the NRMSE for higher $\lambda$ is generally observed, indicating a more robust regression producing models whose prediction accuracy is case dependent in terms of validation/test time series.
However, the effect of increasing $\lambda$ is found to be beneficial for accuracy (lower average NRMSE) only for hyperparameter configurations involving longer training signals and higher numbers of added states and added inputs, which are combinations that are prone to produce unstable models if not regularized, as evidenced in \cite{palma2025si}. The use of large values for $\lambda$ in such cases appears to be a good strategy to unveil the benefit of increased dimensionality of the training data.
For small dimension HDMDc models (small $l_{tr}$, $l_{d_x}$, and $l_{d_u}$), on the contrary, higher $\lambda$ values lead to higher NRMSE. 
Starting from these observations, an automatic tuning procedure for the optimal $\lambda$ based on the values of the other hyperparameters may be explored in future works.

The time-resolved comparison between HDMDc predictions (for the best NRMSE hyperparameters) and EFD in \cref{fig:BHDMDc_1} shows a fair agreement reached by the DMD-based model with the experimental measurements. The dynamics between higher and smaller amplitude oscillations of the motion variables is well captured, but occasionally peak values can be underestimated. 
However, this statement is partial since it is based on a single model, \textit{i.e.,} a single training sequence randomly selected. 
Exploiting the ensemble of multiple models trained with different sequences, the FHDMDc greatly improves the predictions compared to HDMDc, see \cref{fig:FHDMDc_1} and also \cref{fig:detvsuq}, with particular regard to peak values. Moreover, the uncertainty band of its predictions almost always includes the ground truth from the experiments, showing the importance of statistically considering the choice of the training signal. 

The Bayesian strategy for uncertainty quantification, on the contrary, does not show improvements in the average NRMSE of its average prediction compared to the deterministic best.
More critically, it fails to produce an accurate estimation of the confidence interval, thus misleading the user into believing the model is more accurate than it truly is.
In particular, uncertainty is too small, meaning that the change of the hyperparameters does not produce a sufficient variation in the model and its predictions. 
However, other studies \cite{palma2025si} have shown that this is not always true. Future extensions of the framework should thus consider the combination of the BHDMDc and FHDMDc, for a complete assessment of uncertainty, \textit{i.e.,} coming from both the training sequence and the hyperparameters' settings.

The reproduction accuracy of HDMDc, BHDMDc, and FHDMDc appears to be very different between motion variables and resistance force. This is evident from a qualitative assessment of \cref{fig:BHDMDc_1,fig:FHDMDc_1}, and comparing the PDFs from the MBB analysis in \cref{fig:mbb}: there is an order of magnitude of difference in the JSD evaluated for, \textit{e.g.,} $\ddot{z}_B$ and $F_x$, see \cref{tab:mbbjsd}. 
The test time series comparison shows that the DMD-based models are able to fairly reproduce the main low-frequency component of the total resistance in waves, however, filtering out higher harmonics. It shall be noted how the FHDMDc uncertainty almost always includes the EFD prediction also for this variable, demonstrating notable robustness.
In order to improve the prediction capabilities of resistance, future works shall explore the possibility of building two distinct models for motions and forces, also in a two-step fashion, with the force model exploiting the predicted motions as inputs. 
In addition, methodological extensions such as the extended DMD, combined with HDMDc and its ensemble variants, shall be considered to include stronger nonlinear input-output relations in the DMD regression by means of nonlinear observables of the inputs and/or state variables.

A final remark is dedicated to the comparison of the PDFs on boostrapped time series from FHDMDc and CFD in \cref{fig:mbb}. 
The JSD of the PDFs of the two numerical methods compared to EFD is overall comparable for motion variables, and their average lines always fall within the experimental uncertainty band.
In addition, a qualitative evaluation of the PDFs shapes highlights that trends can be even better captured by FHDMDc for some variables, see the slopes and peak location in $-30\le z(\text{mm}) \le 50$, $3 \le \theta(\text{deg}) \le 5$ and $\dot{z}_D$.

\section{Conclusions}\label{s:conclusion}

In this study, an ensemble-based Hankel Dynamic Mode Decomposition with control (HDMDc) framework was introduced and validated for uncertainty-aware seakeeping prediction of the Delft 372 catamaran. 
Analyses have been conducted using experimental data from captive seakeeping experiments in irregular waves conducted in the CNR-INM facility.

By augmenting the classical DMDc formulation with time-lagged state and input snapshots and deploying two distinct ensembling strategies, namely Bayesian HDMDc (BHDMDc) and Frequentist HDMDc (FHDMDc), both mean forecasts and confidence intervals were produced for key responses identified as per the NATO STANAG 4154 standardization agreement (heave, pitch, flight-deck velocity, bridge acceleration, and total resistance) under irregular head waves.

Frequentist ensembling (FHDMDc) markedly improved predictive accuracy over deterministic HDMDc, particularly in capturing peak motions; its uncertainty bands consistently encompass experimental measurements across all degrees of freedom.

Bayesian ensembling (BHDMDc), despite conceptually promising, did not yield accuracy gains in this application and tended to under-estimate predictive uncertainty, suggesting that hyperparameter variability alone cannot span the true model error.

The regularization parameter $\lambda$ proves critical for stabilizing HDMDc regression when using large augmented dimensions: higher $\lambda$ values reduce error dispersion for long training windows and multiple delays, but excessive regularization can degrade performance in low-dimensional settings.

A statistical comparison of probability density functions via moving-block bootstrap and Jensen–Shannon divergence further demonstrates that FHDMDc substantially matches high-fidelity URANS simulations in reproducing full response distributions. The low computational footprint of the FHDMDc method positions it as a strong candidate for operational forecasting.

Despite the strong performance in motion prediction, resistance forecasts still exhibit attenuation of high-frequency content. Future efforts will explore two-stage architectures (motions feeding resistance prediction), and the incorporation of extended DMD observables to capture nonlinear input–output coupling. Additionally, a hybrid uncertainty framework combining training-set variability with hyperparameter uncertainty could provide more comprehensive confidence assessments.

In summary, the ensemble HDMDc approach delivers a compact, interpretable, and computationally efficient tool for seakeeping prediction with quantified uncertainty, bridging the gap between high-fidelity simulations and black-box machine-learning methods for marine design and operational support.

Funding
Conflict of Interest
Author contributions (if more than 1 author)
Data Availability
Ethics approval and Consent to participate
Acknowledgments

\section*{Funding}
This work is supported by the US Office of Naval Research Global, grant N62909-24-1-2102, under the administration of Drs. Richard Meyer and Robert Brizzolara, and 
by the Italian Ministry of University and Research through the National Recovery and Resilience Plan (PNRR), CN00000023 - CUP B43C22000440001, “Sustainable Mobility Center” (CNMS), Spoke 3 “Waterways”.

\section*{Conflict of interest}
There are no financial or personal relationships that could have appeared to influence the work reported in this paper.

\section*{CRediT authorship contribution statement}
\textbf{Giorgio Palma:} Conceptualization, Methodology, Software, Validation, Investigation, Formal Analysis, Writing - Original Draft, Visualization.
\textbf{Andrea Serani:} Conceptualization, Methodology, Resources, Writing - Review \& Editing.
\textbf{Matteo Diez:} Conceptualization, Methodology, Investigation, Data Curation, Resources, Writing - Review \& Editing, Supervision, Funding acquisition.

\section*{Acknowledgments}
Drs. Danilo Durante and Riccardo Broglia at CNR-INM are gratefully acknowledged for providing the experimental data used in the present work.



\bibliographystyle{unsrt}  
\bibliography{biblio}  

@inproceedings{stern2022kcs,
  author    = {Stern, F. and Sanada, Y. and Park, S. and Wang, Z. and Yasukawa, H. and Diez, M. and Quadvlieg, F. and Bedos, A.},
  title     = {Experimental and CFD Study of KCS Turning Circles in Waves},
  booktitle = {Proceedings of the 34th Symposium on Naval Hydrodynamics},
  year      = {2022},
  month     = {6},
  address   = {Washington DC, USA}
}

@article{Sanada2021,
    author = {Sanada, Yugo  and Park, Sungtek and Kim, Dong-Hwan and Wang, Zhaoyuan and Stern, Frederick and Yasukawa, Hironori},
    title = {Experimental and computational study of hull–propeller–rudder interaction for steady turning circles},
    journal = {Physics of Fluids},
    volume = {33},
    number = {12},
    pages = {127117},
    year = {2021},
    month = {12},
    issn = {1070-6631},
    doi = {10.1063/5.0073098},
    url = {https://doi.org/10.1063/5.0073098},
    eprint = {https://pubs.aip.org/aip/pof/article-pdf/doi/10.1063/5.0073098/16751782/127117\_1\_online.pdf},
}

@article{Yasukawa2016,
    author = {Yasukawa, Hironori and Hirata, Noritaka and Nakayama, Yoshiyuki},
    title = {High-Speed Ship Maneuverability},
    journal = {Journal of Ship Research},
    volume = {60},
    number = {04},
    pages = {239-258},
    year = {2016},
    month = {12},
    issn = {0022-4502},
    doi = {10.5957/jsr.2016.60.4.239},
    url = {https://doi.org/10.5957/jsr.2016.60.4.239},
    eprint = {https://onepetro.org/JSR/article-pdf/60/04/239/2232388/sname-jsr-2016-60-4-239.pdf},
}

@Article{Yasukawa2015,
author={Yasukawa, H.
and Yoshimura, Y.},
title={Introduction of MMG standard method for ship maneuvering predictions},
journal={Journal of Marine Science and Technology},
year={2015},
month={3},
day={01},
volume={20},
number={1},
pages={37-52},
issn={1437-8213},
doi={10.1007/s00773-014-0293-y},
url={https://doi.org/10.1007/s00773-014-0293-y}
}

@article{palma2025si,
title = {Model-free system identification of surface ships in waves via Hankel dynamic mode decomposition with control},
journal = {Ocean Engineering},
volume = {341},
pages = {122539},
year = {2025},
issn = {0029-8018},
NOOPdoi = {https://doi.org/10.1016/j.oceaneng.2025.122539},
NOOPurl = {https://www.sciencedirect.com/science/article/pii/S002980182502222X},
author = {Giorgio Palma and Andrea Serani and Shawn Aram and David W. Wundrow and David Drazen and Matteo Diez},
}

@article{palma2024forecasting,
title = {Bayesian Hankel dynamic mode decomposition for ship motion digital twinning},
journal = {Applied Ocean Research},
volume = {165},
pages = {104863},
year = {2025},
issn = {0141-1187},
doi = {https://doi.org/10.1016/j.apor.2025.104863},
url = {https://www.sciencedirect.com/science/article/pii/S0141118725004481},
author = {Giorgio Palma and Andrea Serani and Shawn Aram and David W. Wundrow and David Drazen and Matteo Diez},
}

@Article{palma2025windturbine,
AUTHOR = {Palma, Giorgio and Bardazzi, Andrea and Lucarelli, Alessia and Pilloton, Chiara and Serani, Andrea and Lugni, Claudio and Diez, Matteo},
TITLE = {Analysis, Forecasting, and System Identification of a Floating Offshore Wind Turbine Using Dynamic Mode Decomposition},
JOURNAL = {Journal of Marine Science and Engineering},
VOLUME = {13},
YEAR = {2025},
MONTH = {3},
NUMBER = {4},
ARTICLE-NUMBER = {656},
URL = {https://www.mdpi.com/2077-1312/13/4/656},
ISSN = {2077-1312},
DOI = {10.3390/jmse13040656}
}

@inproceedings{diez2024sname,
    author = {Diez, Matteo  and Wang, Zhaoyuan  and Park, Sungtek  and Milano, Christian  and Stern, Frederick  and Yasukawa, Hironori  and Gunderson, Andrew  and Scherer, John },
    title = {Multi-Fidelity MMG-Model for Digital Design of High-Speed Small Craft},
    booktitle = {SNAME Chesapeake Power Boat Symposium},    
    pages = {D011S002R003},
    year = {2024},
    month = {10},
    doi = {10.5957/CPBS-2024-008},
    NOOPurl = {https://doi.org/10.5957/CPBS-2024-008},
    NOOPeprint = {https://onepetro.org/snamecpbs/proceedings-pdf/CPBS24/CPBS24/D011S002R003/3683932/sname-cpbs-2024-008.pdf},
}

@article{Brunton2021,
author = {Brunton, Steven L. and Budi\v{s}i\'{c}, Marko and Kaiser, Eurika and Kutz, J. Nathan},
title = {Modern Koopman Theory for Dynamical Systems},
journal = {SIAM Review},
volume = {64},
number = {2},
pages = {229-340},
year = {2022},
doi = {10.1137/21M1401243},
noopURL = {https://doi.org/10.1137/21M1401243},
noopeprint = {https://doi.org/10.1137/21M1401243},
}

@article{mezic2017,
author = {Arbabi, Hassan and Mezi\'{c}, Igor},
title = {Ergodic Theory, Dynamic Mode Decomposition, and Computation of Spectral Properties of the Koopman Operator},
journal = {SIAM Journal on Applied Dynamical Systems},
volume = {16},
number = {4},
pages = {2096-2126},
year = {2017},
doi = {10.1137/17M1125236},
NOOPurl = {        https://doi.org/10.1137/17M1125236},
NOOPeprint = {         https://doi.org/10.1137/17M1125236}
}

@article{brunton2016b,
    doi = {10.1371/journal.pone.0150171},
    author = {Brunton, Steven L. AND Brunton, Bingni W. AND Proctor, Joshua L. AND Kutz, J. Nathan},
    journal = {PLOS ONE},
    publisher = {Public Library of Science},
    title = {{Koopman} Invariant Subspaces and Finite Linear Representations of Nonlinear Dynamical Systems for Control},
    year = {2016},
    month = {2},
    volume = {11},
    NOOPurl = {https://doi.org/10.1371/journal.pone.0150171},
    pages = {1-19},
    number = {2},
}

@book{kutz2016dynamic,
author = {Kutz, J. and Brunton, Steven and Brunton, Bingni and Proctor, Joshua},
title = {Dynamic Mode Decomposition: Data-Driven Modeling of Complex Systems},
year = {2016},
month = {11},
pages = {},
publisher = {SIAM - Society for Industrial and Applied Mathematics},
isbn = {978-1-611974-49-2}
}

@article{Tu2014,
author = {Tu, Jonathan H. and Rowley, Clarence W. and Luchtenburg, Dirk M. and Brunton, Steven L. and Kutz, J. Nathan},
title = {On dynamic mode decomposition:  Theory and applications},
journal = {Journal of Computational Dynamics},
volume = {1},
number = {2},
pages = {391-421},
year = {2014},
issn = {2158-2491},
doi = {10.3934/jcd.2014.1.391},
noopURL = {https://www.aimsciences.org/article/id/1dfebc20-876d-4da7-8034-7cd3c7ae1161}
}

@Article{Brunton2017,
author={Brunton, Steven L.
and Brunton, Bingni W.
and Proctor, Joshua L.
and Kaiser, Eurika
and Kutz, J. Nathan},
title={Chaos as an intermittently forced linear system},
journal={Nature Communications},
year={2017},
month={5},
day={30},
volume={8},
number={1},
pages={19},
issn={2041-1723},
doi={10.1038/s41467-017-00030-8},
noopURL={https://doi.org/10.1038/s41467-017-00030-8}
}

@article{marlantes2024,
title = {Predicting ship responses in different seaways using a generalizable force correcting machine learning method},
journal = {Ocean Engineering},
volume = {312},
pages = {119110},
year = {2024},
issn = {0029-8018},
doi = {https://doi.org/10.1016/j.oceaneng.2024.119110},
noopURL = {https://www.sciencedirect.com/science/article/pii/S002980182402448X},
author = {Marlantes, Kyle E. and Bandyk, Piotr J. and Maki, Kevin J.}
}

@ARTICLE{Lin1991,
  author={Lin, J.},
  journal={IEEE Transactions on Information Theory}, 
  title={Divergence measures based on the Shannon entropy}, 
  year={1991},
  volume={37},
  number={1},
  pages={145-151},
  doi={10.1109/18.61115}
}

@article{Kullback1951,
 ISSN = {00034851},
 noopURL = {http://www.jstor.org/stable/2236703},
 author = {Kullback, S. and Leibler, R. A.},
 journal = {The Annals of Mathematical Statistics},
 number = {1},
 pages = {79--86},
 publisher = {Institute of Mathematical Statistics},
 title = {On Information and Sufficiency},
 urldate = {2024-05-22},
 volume = {22},
 year = {1951}
}

@article{Marusic2024, 
title={Dynamic mode decomposition for analysis of time-series data}, 
volume={1000}, 
DOI={10.1017/jfm.2024.834}, 
journal={Journal of Fluid Mechanics}, 
author={Marusic, Ivan}, 
year={2024}, 
pages={F7}}

@article{Koopman1931,
author = {Koopman, B. O.},
title = {Hamiltonian Systems and Transformation in Hilbert Space},
journal = {Proceedings of the National Academy of Sciences},
volume = {17},
number = {5},
pages = {315-318},
year = {1931},
doi = {10.1073/pnas.17.5.315},
NOOPurl = {https://www.pnas.org/doi/abs/10.1073/pnas.17.5.315},
NOOPeprint = {https://www.pnas.org/doi/pdf/10.1073/pnas.17.5.315}}

@article{schmid2010,
  title        = {Dynamic mode decomposition of numerical and experimental data},
  author       = {Schmid, PETER J.},
  year         = 2010,
  journal      = {Journal of Fluid Mechanics},
  publisher    = {Cambridge University Press},
  volume       = 656,
  pages        = {5–28},
  doi          = {10.1017/S0022112010001217}
}

@article{schmid2008dynamic,
  title={Dynamic Mode Decomposition of Numerical and Experimental Data In: Sixty-First Annual Meeting of the APS Division of Fluid Dynamics},
  author={Schmid, PJ and Sesterhenn, J},
  journal={San Antonio, Texas, USA},
  year={2008}
}

@article{rowley2009, title={Spectral analysis of nonlinear flows}, volume={641}, DOI={10.1017/S0022112009992059}, journal={Journal of Fluid Mechanics}, author={Rowley, Clarence W. and Mezi\'{c}, IGOR and Bagheri, SHERVIN and Schlatter, PHILIPP and Hennigson, DAN S.}, year={2009}, pages={115–127}}

@Article{Semeraro2012,
author={Semeraro, Onofrio
and Bellani, Gabriele
and Lundell, Fredrik},
title={Analysis of time-resolved PIV measurements of a confined turbulent jet using POD and Koopman modes},
journal={Experiments in Fluids},
year={2012},
month={11},
day={01},
volume={53},
number={5},
pages={1203-1220},
issn={1432-1114},
doi={10.1007/s00348-012-1354-9},
noopURL={https://doi.org/10.1007/s00348-012-1354-9}
}

@article{Song2013,
    author = {Song, G. and Alizard, F. and Robinet, J.-C. and Gloerfelt, X.},
    title = "{Global and Koopman modes analysis of sound generation in mixing layers}",
    journal = {Physics of Fluids},
    volume = {25},
    number = {12},
    pages = {124101},
    year = {2013},
    month = {12},
    issn = {1070-6631},
    doi = {10.1063/1.4834438},
    noopURL = {https://doi.org/10.1063/1.4834438},
    noopeprint = {https://pubs.aip.org/aip/pof/article-pdf/doi/10.1063/1.4834438/15943560/124101\_1\_online.pdf},
}

@Article{Tang2012,
author={Tang, ZhanQi
and Jiang, Nan},
title={Dynamic mode decomposition of hairpin vortices generated by a hemisphere protuberance},
journal={Science China Physics, Mechanics and Astronomy},
year={2012},
month={1},
day={01},
volume={55},
number={1},
pages={118-124},
issn={1869-1927},
doi={10.1007/s11433-011-4535-2},
noopURL={https://doi.org/10.1007/s11433-011-4535-2}
}

@article{proctor2016dynamic,
  title        = {Dynamic mode decomposition with control},
  author       = {Proctor, Joshua L and Brunton, Steven L and Kutz, J Nathan},
  year         = {2016},
  volume = {15},
  number = {1},
  journal      = {SIAM Journal on Applied Dynamical Systems},
  publisher    = {SIAM},
  pages        = {142--161}
}

@article{Proctor2015,
    author = {Proctor, Joshua L. and Eckhoff, Philip A.},
    title = "{Discovering dynamic patterns from infectious disease data using dynamic mode decomposition}",
    journal = {International Health},
    volume = {7},
    number = {2},
    pages = {139-145},
    year = {2015},
    month = {02},
    issn = {1876-3413},
    doi = {10.1093/inthealth/ihv009},
    noopURL = {https://doi.org/10.1093/inthealth/ihv009},
    noopeprint = {https://academic.oup.com/inthealth/article-pdf/7/2/139/16837591/ihv009.pdf},
}

@article{brunton2016,
title = {Extracting spatial–temporal coherent patterns in large-scale neural recordings using dynamic mode decomposition},
journal = {Journal of Neuroscience Methods},
volume = {258},
pages = {1-15},
year = {2016},
issn = {0165-0270},
doi = {https://doi.org/10.1016/j.jneumeth.2015.10.010},
noopURL = {https://www.sciencedirect.com/science/article/pii/S0165027015003829},
author = {Brunton, Bingni W. and Johnson, Lise A. and Ojemann, Jeffrey G. and Kutz, J. Nathan}
}

@article{mann2016,
author = {Mann, Jordan and Kutz, J. Nathan},
title = {Dynamic mode decomposition for financial trading strategies},
journal = {Quantitative Finance},
volume = {16},
number = {11},
pages = {1643--1655},
year = {2016},
publisher = {Routledge},
doi = {10.1080/14697688.2016.1170194},
noopURL = { https://doi.org/10.1080/14697688.2016.1170194},
noopeprint = { https://doi.org/10.1080/14697688.2016.1170194}
}

@Article{diez2022datadriven,
author={Diez, Matteo
and Serani, Andrea
and Campana, Emilio F.
and Stern, Frederick},
title={Time-series forecasting of ships maneuvering in waves via dynamic mode decomposition},
journal={Journal of Ocean Engineering and Marine Energy},
year={2022},
month={11},
day={01},
volume={8},
number={4},
pages={471-478},
issn={2198-6452},
doi={10.1007/s40722-022-00243-0},
noopURL={https://doi.org/10.1007/s40722-022-00243-0}
}

@inproceedings{diez2022snh,
author = {Diez, Matteo and Serani, Andrea and Gaggero, Mauro and Campana, Emilio Fortunato},
title = {Improving Knowledge and Forecasting of Ship Performance in Waves via Hybrid Machine Learning Methods},
booktitle = {34th Symposium on Naval Hydrodynamics, Washington, DC, USA, June 26 - July 1, 2022},
year = {2022}
}

@article{Diez2024,
author = {Diez, Matteo and Gaggero, Mauro and Serani, Andrea},
title = {Data-driven forecasting of ship motions in waves using machine learning and dynamic mode decomposition},
journal = {International Journal of Adaptive Control and Signal Processing},
year = {2024},
volume = {n/a},
number = {n/a},
pages = {},
doi = {https://doi.org/10.1002/acs.3835},
noopURL = {https://onlinelibrary.wiley.com/doi/abs/10.1002/acs.3835},
noopeprint = {https://onlinelibrary.wiley.com/doi/pdf/10.1002/acs.3835}
}

@article{serani2023,
title = {On the use of dynamic mode decomposition for time-series forecasting of ships operating in waves},
journal = {Ocean Engineering},
volume = {267},
pages = {113235},
year = {2023},
issn = {0029-8018},
doi = {https://doi.org/10.1016/j.oceaneng.2022.113235},
noopURL = {https://www.sciencedirect.com/science/article/pii/S0029801822025185},
author = {Serani, Andrea and Dragone, Paolo and Stern, Frederick and Diez, Matteo}
}

@article{serani2021urans,
  title        = {{URANS} analysis of a free-running destroyer sailing in irregular stern-quartering waves at sea state 7},
  author       = {Serani, Andrea and Diez, Matteo and van Walree, Frans and Stern, Frederick},
  year         = 2021,
  journal      = {Ocean Engineering},
  publisher    = {Elsevier},
  volume       = 237,
  pages        = 109600
}

@article{Miecznikowski2010,
author = {Miecznikowski, Jeffrey C. and Wang, Dongliang and Hutson, Alan},
title = {Bootstrap MISE Estimators to Obtain Bandwidth for Kernel Density Estimation},
journal = {Communications in Statistics - Simulation and Computation},
volume = {39},
number = {7},
pages = {1455--1469},
year = {2010},
publisher = {Taylor \& Francis},
doi = {10.1080/03610918.2010.500108},
URL = { https://doi.org/10.1080/03610918.2010.500108},
eprint = {https://doi.org/10.1080/03610918.2010.500108}
}

@book{Silverman2018,
  title={Density estimation for statistics and data analysis},
  author={Silverman, Bernard W},
  year={2018},
  publisher={Routledge},
  doi={https://doi.org/10.1201/9781315140919}
}

@article{chen2023,
    author = {Chen, Chang-Zhe and Zou, Zao-Jian and Zou, Lu and Zou, Ming and Kou, Jia-Qing},
    title = {Time series prediction of ship course keeping in waves using higher order dynamic mode decomposition},
    journal = {Physics of Fluids},
    volume = {35},
    number = {9},
    pages = {097139},
    year = {2023},
    month = {09},
    issn = {1070-6631},
    doi = {10.1063/5.0165665},
    url = {https://doi.org/10.1063/5.0165665},
    eprint = {https://pubs.aip.org/aip/pof/article-pdf/doi/10.1063/5.0165665/18135297/097139\_1\_5.0165665.pdf},
}

@article{chen2024,
title = {Real-time prediction of ship maneuvering motion in waves based on an improved reduced-order model},
journal = {Ocean Engineering},
volume = {312},
pages = {119244},
year = {2024},
issn = {0029-8018},
doi = {https://doi.org/10.1016/j.oceaneng.2024.119244},
url = {https://www.sciencedirect.com/science/article/pii/S0029801824025824},
author = {Chen, Chang-Zhe and Liu, Si-Yu and Zou, Zao-Jian and Zou, Lu}
}

@Article{mezic2022,
AUTHOR = {Mezić, Igor},
TITLE = {On Numerical Approximations of the {Koopman} Operator},
JOURNAL = {Mathematics},
VOLUME = {10},
YEAR = {2022},
NUMBER = {7},
ARTICLE-NUMBER = {1180},
NOOPurl = {https://www.mdpi.com/2227-7390/10/7/1180},
ISSN = {2227-7390},
DOI = {10.3390/math10071180}
}

@inproceedings{Serani2024snh,
  author    = {Serani, A. and Diez, M. and Aram, S. and Wundrow, D. and Drazen, D. and McTaggart, K.},
  title     = {Model Order Reduction of 5415M in Irregular Waves via Dynamic Mode Decomposition: Computational Models’ Diagnostics, Forecasting, and System Identification Capabilities},
  booktitle = {Proceedings of the 35th Symposium on Naval Hydrodynamics (SNH)},
  year      = {2024},
  month     = {07},
  address   = {Nantes, France},
  note      = {Held July 7--13, 2024},
}

@article{XIE2024,
title = {Regularized dynamic mode decomposition algorithm for time sequence predictions},
journal = {Theoretical and Applied Mechanics Letters},
volume = {14},
number = {5},
pages = {100555},
year = {2024},
issn = {2095-0349},
doi = {https://doi.org/10.1016/j.taml.2024.100555},
url = {https://www.sciencedirect.com/science/article/pii/S2095034924000667},
author = {Xiaoyang Xie and Shaoqiang Tang}
}

@book{vantveer1998,
title = "Experimental results of motions, hydrodynamic coefficients and wave loads on the 372 catamaran model",
author = "{van 't Veer}, A. P.",
note = "1129",
year = "1998",
publisher = "WbMT",
}

@article{DURANTE2020,
title = {Accurate experimental benchmark study of a catamaran in regular and irregular head waves including uncertainty quantification},
journal = {Ocean Engineering},
volume = {195},
pages = {106685},
year = {2020},
issn = {0029-8018},
doi = {https://doi.org/10.1016/j.oceaneng.2019.106685},
url = {https://www.sciencedirect.com/science/article/pii/S0029801819308005},
author = {D. Durante and R. Broglia and M. Diez and A. Olivieri and E.F. Campana and F. Stern}
}

@inproceedings{PANDEY2016,
title = {Study on Turning Manoeuvre of Catamaran Surface Vessel with a Combined Experimental and Simulation Method},
volume = {49},
number = {23},
pages = {446-451},
year = {2016},
booktitle = {10th IFAC Conference on Control Applications in Marine Systems CAMS 13—16 September 2016},
issn = {2405-8963},
doi = {https://doi.org/10.1016/j.ifacol.2016.10.446},
url = {https://www.sciencedirect.com/science/article/pii/S2405896316320341},
author = {Pandey, Jyotsna and Hasegawa, Kazuhiko},
address={Trondheim, Norway}
}

@inproceedings{pandey2016manoeuvring,
  title={Manoeuvring mathematical model of catamaran wave adaptive modular vessel (WAM-V) using the system identification technique},
  author={Pandey, Jyotsna and Hasegawa, Kazuhiko},
  booktitle={Proceedings of 7th PAAMES and AMEC2016},
  volume={13},
  pages={14},
  year={2016},
  address={Hong Kong, China}
}

@article{castiglione2011,
title = {Numerical investigation of the seakeeping behavior of a catamaran advancing in regular head waves},
journal = {Ocean Engineering},
volume = {38},
number = {16},
pages = {1806-1822},
year = {2011},
issn = {0029-8018},
doi = {https://doi.org/10.1016/j.oceaneng.2011.09.003},
url = {https://www.sciencedirect.com/science/article/pii/S0029801811001971},
author = {Teresa Castiglione and Frederick Stern and Sergio Bova and Manivannan Kandasamy}
}

@article{broglia2019,
    author = {Broglia, R. and Zaghi, S. and Campana, E. F. and Dogan, T. and Sadat-Hosseini, H. and Stern, F. and Queutey, P. and Visonneau, M. and Milanov, E.},
    title = {Assessment of Computational Fluid Dynamics Capabilities for the Prediction of Three-Dimensional Separated Flows: The DELFT 372 Catamaran in Static Drift Conditions},
    journal = {Journal of Fluids Engineering},
    volume = {141},
    number = {9},
    pages = {091105},
    year = {2019},
    month = {03},
    issn = {0098-2202},
    doi = {10.1115/1.4042752},
    url = {https://doi.org/10.1115/1.4042752},
    eprint = {https://asmedigitalcollection.asme.org/fluidsengineering/article-pdf/141/9/091105/6400122/fe\_141\_09\_091105.pdf},
}

@inproceedings{broglia2015cfd,
  title={CFD validation for DELFT 372 catamaran in static drift conditions, including onset and progression analysis},
  author={Broglia, Riccardo and Zaghi, Stefano and Campana, Emilio F and Visonneau, Michel and Queutey, Patrick and Dogan, Timur and Sadat-Hosseini, H and Stern, Frederik and Milanov, E},
  booktitle={SNAME Maritime Convention},
  pages={D021S006R014},
  year={2015},
  organization={SNAME}
}

@inproceedings{sadat2013cfd,
  title={CFD and system-based prediction of Delft catamaran maneuvering in calm water and regular waves},
  author={Sadat-Hosseini, H and Chen, X and Milanov, E and Stern, F},
  booktitle={Proc. 12th International Conference on Fast Sea Transportation, FAST},
  pages={2--5},
  year={2013}
}

@article{silva2022,
title = {Data-Driven system identification of 6-DoF ship motion in waves with neural networks},
journal = {Applied Ocean Research},
volume = {125},
pages = {103222},
year = {2022},
issn = {0141-1187},
doi = {https://doi.org/10.1016/j.apor.2022.103222},
url = {https://www.sciencedirect.com/science/article/pii/S0141118722001614},
author = {Silva, Kevin M. and Maki, Kevin J.}
}

@inproceedings{diez2013reliability,
  title={Reliability-based robust design optimization for ships in real ocean environment},
  author={Diez, M and Chen, X and Campana, EF and Stern, F},
  booktitle={12th International Conference on Fast Sea Transportation, FAST2013, Amsterdam, The Netherlands},
  year={2013}
}

@CONFERENCE{Kennell1985,
	author = {Kennell, Colen G. and White, Brian L. and Comstock, Edward N.},
	title = {INNOVATIVE NAVAL DESIGNS FOR NORTH ATLANTIC OPERATIONS.},
	year = {1985},
	journal = {Transactions - Society of Naval Architects and Marine Engineers},
	volume = {93},
	pages = {261 – 281},
	url = {https://www.scopus.com/inward/record.uri?eid=2-s2.0-0022225296&partnerID=40&md5=87bace8b1b5a2b7ae6a66cc9a22db42c},
	publisher = {SNAME}
}

@misc{STANAG4154,
    author = {{NATO STANAG 4154}},
    title = {Common Procedures in the Ship Design Process. Seakeeping Criteria for General Application (Chapter 7)},
    year = {1997},
}

@article{carlstein1986use,
  title={The use of subseries values for estimating the variance of a general statistic from a stationary sequence},
  author={Carlstein, Edward},
  journal={The annals of statistics},
  pages={1171--1179},
  year={1986},
  publisher={JSTOR}
}

@article{diez2018statistical,
    author = {Diez, Matteo and Broglia, Riccardo and Durante, Danilo and Olivieri, Angelo and Campana, Emilio F. and Stern, Frederick},
    title = {Statistical Assessment and Validation of Experimental and Computational Ship Response in Irregular Waves},
    journal = {Journal of Verification, Validation and Uncertainty Quantification},
    volume = {3},
    number = {2},
    pages = {021004},
    year = {2018},
    month = {10},
    issn = {2377-2158},
    doi = {10.1115/1.4041372},
    url = {https://doi.org/10.1115/1.4041372},
    eprint = {https://asmedigitalcollection.asme.org/verification/article-pdf/3/2/021004/6381759/vvuq\_003\_02\_021004.pdf},
}

@article{aram2024cfd,
  title        = {{CFD} validation and analysis of turning maneuvers of a surface combatant in regular waves},
  author       = {Aram, Shawn and Mucha, Philipp},
  year         = 2024,
  journal      = {Ocean Engineering},
  publisher    = {Elsevier},
  volume       = 293,
  pages        = 116653
}

@Article{DAgostino2022,
author={D'Agostino, Danny
and Serani, Andrea
and Stern, Frederick
and Diez, Matteo},
title={Time-series forecasting for ships maneuvering in waves via recurrent-type neural networks},
journal={Journal of Ocean Engineering and Marine Energy},
year={2022},
month={11},
day={01},
volume={8},
number={4},
pages={479-487},
issn={2198-6452},
doi={10.1007/s40722-022-00255-w},
noopurl={https://doi.org/10.1007/s40722-022-00255-w}
}

@article{XU2021,
title = {A data-driven model for nonlinear marine dynamics},
journal = {Ocean Engineering},
volume = {236},
pages = {109469},
year = {2021},
issn = {0029-8018},
doi = {https://doi.org/10.1016/j.oceaneng.2021.109469},
url = {https://www.sciencedirect.com/science/article/pii/S0029801821008726},
author = {Xu, Wenzhe and Maki, Kevin J. and Silva, Kevin M.}
}

@article{Wang2023,
title = {SeaBil: Self-attention-weighted ultrashort-term deep learning prediction of ship maneuvering motion},
journal = {Ocean Engineering},
volume = {287},
pages = {115890},
year = {2023},
issn = {0029-8018},
doi = {https://doi.org/10.1016/j.oceaneng.2023.115890},
noopURL = {https://www.sciencedirect.com/science/article/pii/S0029801823022746},
author = {Wang, Ning and Kong, Xiangjun and Ren, Boyu and Hao, Lizhu and Han, Bing}
}

@Article{jiang2024,
AUTHOR = {Jiang, Zhiqiang and Ma, Yongyan and Li, Weijia},
TITLE = {A Data-Driven Method for Ship Motion Forecast},
JOURNAL = {Journal of Marine Science and Engineering},
VOLUME = {12},
YEAR = {2024},
NUMBER = {2},
ARTICLE-NUMBER = {291},
noopURL = {https://www.mdpi.com/2077-1312/12/2/291},
ISSN = {2077-1312},
DOI = {10.3390/jmse12020291}
}

@inproceedings{takeishi2017b,
  author    = {Takeishi, Naoya and Kawahara, Yoshinobu and Tabei, Yasuo and Yairi, Takehisa},
  title     = {Bayesian Dynamic Mode Decomposition},
  booktitle = {Proceedings of the Twenty-Sixth International Joint Conference on
               Artificial Intelligence, {IJCAI-17}},
  pages     = {2814--2821},
  year      = {2017},
  doi       = {10.24963/ijcai.2017/392},
  noopURL   = {https://doi.org/10.24963/ijcai.2017/392},
}

@article{Sashidhar2022,
author = {Sashidhar, Diya  and Kutz, J. Nathan },
title = {Bagging, optimized dynamic mode decomposition for robust, stable forecasting with spatial and temporal uncertainty quantification},
journal = {Philosophical Transactions of the Royal Society A: Mathematical, Physical and Engineering Sciences},
volume = {380},
number = {2229},
pages = {20210199},
year = {2022},
doi = {10.1098/rsta.2021.0199},
noopURL = {https://royalsocietypublishing.org/doi/abs/10.1098/rsta.2021.0199},
noopeprint = {https://royalsocietypublishing.org/doi/pdf/10.1098/rsta.2021.0199}
}

@article{askham2018,
author = {Askham, Travis and Kutz, J. Nathan},
title = {Variable Projection Methods for an Optimized Dynamic Mode Decomposition},
journal = {SIAM Journal on Applied Dynamical Systems},
volume = {17},
number = {1},
pages = {380-416},
year = {2018},
doi = {10.1137/M1124176},
noopURL = {https://doi.org/10.1137/M1124176},
noopEPRINT = {https://doi.org/10.1137/M1124176}
}

\appendix
\end{document}